\documentclass[aps,prx,reprint,superscriptaddress,twocolumn,floatfix]{revtex4}

\usepackage{graphicx} 
\usepackage{amsmath,mathtools,amsfonts,amssymb}
\usepackage{braket}
\usepackage{tikz}
\usepackage{pgfplots}
\usepackage[normalem]{ulem}

\pgfplotsset{compat=newest}

\newcommand{\hatn}{ }

\begin{document}

\title{Thermalization induced by quantum scattering}

\author{Samuel L. Jacob}
\email{samuel.lourenco@uni.lu}
\affiliation{Complex Systems and Statistical Mechanics, Physics and Materials Science Research Unit, University of Luxembourg, L-1511 Luxembourg, G.D. Luxembourg}
\affiliation{Kavli Institute for Theoretical Physics, University of California, Santa Barbara, CA 93106 Santa Barbara,  U.S.A.}

\author{Massimiliano Esposito}
\email{massimiliano.esposito@uni.lu}
\affiliation{Complex Systems and Statistical Mechanics, Physics and Materials Science Research Unit, University of Luxembourg, L-1511 Luxembourg, G.D. Luxembourg}
\affiliation{Kavli Institute for Theoretical Physics, University of California, Santa Barbara, CA 93106 Santa Barbara,  U.S.A.}

\author{Juan M. R. Parrondo}
\email{parrondo@fis.ucm.es}
\affiliation{Departamento de Estructura de la Materia, F\'isica T\'ermica y Electr\'onica and  GISC, Universidad Complutense de Madrid, 28040 Madrid, Spain}

\author{Felipe Barra}
\email{fbarra@dfi.uchile.cl }
\affiliation{Departamento de F\'isica, Facultad de Ciencias F\'isicas y Matem\'aticas, Universidad de Chile, 837.0415 Santiago, Chile}
\affiliation{Kavli Institute for Theoretical Physics, University of California, Santa Barbara, CA 93106 Santa Barbara,  U.S.A.}

\date{\today}



\begin{abstract}
We use quantum scattering theory to study a fixed quantum system $Y$ subject to collisions with massive particles $X$ described by wave-packets. We derive the scattering map for system $Y$ and show that the induced evolution crucially depends on the width of the incident wave-packets compared to the level spacing in $Y$. If $Y$ is non-degenerate, sequential collisions with narrow wave-packets cause $Y$ to decohere. Moreover, an ensemble of narrow packets produced by thermal effusion causes $Y$ to thermalize. On the other hand, broad wave-packets can act as a source of coherences for $Y$, even in the case of an ensemble of incident wave-packets given by the effusion distribution, preventing  thermalization. We illustrate our findings on several simple examples and discuss the consequences of our results in realistic experimental situations.
\end{abstract}
\pacs{}

\maketitle

\section{Introduction}

Many phenomena in quantum physics can be described in terms of repeated collisions of moving particles $X$ with a fixed target system $Y$.
The effect of such collisions on $Y$ is the subject of this paper.
Such a framework is extremely rich, as the particles $X$ can be prepared in an arbitrary state and used to control the dynamics of $Y$.
Without loss of generality, we assume that these particles $X$ have no internal structure. Indeed, we will see that the internal structure of $X$, if decorrelated from its center of mass motion, can be effectively incorporated into that of $Y$.
This allow us to focus on how the motion of the center of mass of such particles ---described by wave-packets--- affects the dynamics of $Y$. 
Besides a notable exception \cite{Englert1991}, it is surprising that such a simple question seems to have been little explored.
For instance, one would expect that phenomena such as decoherence and thermalization of $Y$ occur as a result of collisions with a thermal ensemble of wave-packets $X$, but the conditions for this to happen have not been identified and the precise definition of the thermal ensemble is also lacking.
Our approach contrasts with the theory of open quantum systems, in which a system is in permanent contact with equilibrium reservoirs \cite{Spohn1978,Breuer2007} or subjected to continuous collisions with a quantum gas in thermal equilibrium \cite{Filippov2020}, where such phenomena are well understood.
One would also expect that a non-thermal ensemble, instead of causing decoherence and thermalization, can play the role of a thermodynamic resource for $Y$ and bring it in a suitable coherent state.  
Related questions have been considered within the framework of repeated interaction models \cite{Barra2015,Barra2017,Strasberg2017}, where instead of treating a true scattering problem in real space, an interaction between $X$ and $Y$ is switched on for a given time.

Our results in this paper rely on three main concepts. The first is the scattering map, i.e. the quantum map for $Y$ induced by a collision with a particle $X$ in a generic (pure or mixed) state, defining the reduced dynamics of $Y$. The second is the distinction between pure wave-packets whose energy width is smaller or larger than the smallest energy level spacing in $Y$, which we call narrow or broad wave-packets, respectively. These two concepts suffice to prove our first result: that collisions with narrow wave-packets induce decoherence in a non-degenerate system $Y$, while collisions with broad packets can act as a source of coherences. The third is the notion of a statistical ensemble of effusing narrow packets, which we show not only decoheres but thermalizes $Y$, establishing our second and main result.

This framework offers a rich platform to analyze the interplays between various types of wave-packets for $X$ interacting with $Y$. A quantum thermodynamic perspective on the ensuing free energy transfers is left for future work.

The paper is organized as follows. 
The model and the scattering map for the fixed system $Y$ subject to collisions with wave-packets $X$ are defined in section~\ref{model}. 
The scattering map is then expressed in terms of the scattering matrix in section~\ref{sec.ScatMap}.
Its properties when $X$ is a pure wave-packet are studied in section~\ref{sec.wp}, where we prove that decoherence occurs for collisions with narrow ---but not with broad--- wave-packets. Mixtures, or statistical ensembles, of thermally effusing wave-packets are considered in section~\ref{sec.ensemble}.  
Narrow wave-packets are shown to produce a detailed balance map which leads not only to decoherence of $Y$ but also to thermalization of its populations. Broad wave-packets, on the other hand, can prevent the system from thermalizing by acting as a source of coherences. In section \ref{sec.comment}, we comment on the extension of our framework to particles with an internal structure. In section~\ref{sec.appli}, we illustrate our findings in a simple model based on a Dirac-delta potential.
Conclusions are drawn in section~\ref{sec.conclu}.
Appendix~\ref{sec:appendixscatt} is used to remind some important results from scattering theory used in our calculations. 
The solution of the scattering problem used to illustrate the theory is detailed in Appendix~\ref{app:ejemplo-delta}. 

\section{The model and the scattering map} \label{model}

We consider a particle $X$ with mass $m$ which travels freely before and after colliding with a fixed scatterer, which we call the system $Y$, having internal states in a Hilbert space ${\cal H}_{Y}$. For simplicity, we will restrict ourselves to systems with finite dimension $N$ and a one-dimensional space for the particle. If the latter has no internal structure, the Hilbert space of the global system is ${\cal H}={\cal H}_{X}\otimes {\cal H}_{Y}$, where ${\cal H}_{X}$ is the Hilbert space of a one dimensional particle i.e., the set of normalized wave functions $\psi(x)$ with $x\in {\mathbb R}$ (as we argue in section \ref{sec.conclu}, one can also consider particles with internal structure (e.g. spin) by extending ${\cal H}_{Y}$).

The Hamiltonian for the global system is given by:
	\begin{equation}
	\label{hamiltonianfullYX}
	\hatn{H} = \hatn{H}_{0} + \hatn{V} =  \frac{\hatn{p}^2}{2m} \otimes \hatn{\mathbb{I}}_Y + \hatn{\mathbb{I}}_X \otimes {H}_Y + \hatn{V}(x) \otimes \hatn{\nu}
	\end{equation}
where $x$ and $p$ are the position and momentum operator of the particle $X$. The free Hamiltonian $\hatn{H}_{0} \equiv \frac{\hatn{p}^2}{2m} \otimes \hatn{\mathbb{I}}_Y + \hatn{\mathbb{I}}_X \otimes {H}_Y$ is the sum of the kinetic energy of the particle $X$ and the internal energy of the system $Y$, where $\hatn{\mathbb{I}}_Y$ is the identity in ${\cal H}_{Y}$ (equivalently for $X$) and the interaction is given by $\hatn{V} \equiv \hatn{V}(x) \otimes \hatn{\nu}$,  $\nu$ being an operator in ${\cal H}_{Y}$. Finally, we assume that the interaction potential $V(x)$ tends to zero sufficiently fast as $x \rightarrow \pm \infty$, as is usual in scattering theory \cite{Taylor2006}. 

For such a class of potentials, scattering theory guarantees the existence of a one-to-one map $\hatn{S}$ between free (incoming) states before the collision to free (outgoing) states after the collision. More specifically, the total Hilbert space ${\cal H}$ can be decomposed into the direct sum of scattering and bound states. Only the asymptotic behavior of the former is that of free states, i.e. its evolution for $t\to \pm\infty$ is given by the unitary evolution operator $\hatn{U}_0(t)= \exp[-it \hatn{H}_{0} / \hbar]$ corresponding to the free Hamiltonian $\hatn{H}_{0}$, while the latter remain bound to the interaction region in the same limit. If $\hatn{U}(t) = \exp[-it \hatn{H} / \hbar]$ is the full evolution operator, one can define the isometric M{\o}ller operators \cite{Taylor2006}:
	\begin{equation}
	\label{mollerop}
	\hatn{\Omega}_{\pm} = \lim_{t \rightarrow \mp\infty} \hatn{U}(t)^{\dagger}\hatn{U}_0(t) 
	\end{equation}
which, respectively, map incoming and outgoing states onto scattering states. The scattering operator defined as
	\begin{equation}
	\label{scattop}
	\hatn{S} = \hatn{\Omega}_{-}^{\dagger}\hatn{\Omega}_{+} \; ,
	\end{equation}
is then unitary $\hatn{S} \hatn{S}^{\dagger} = \hatn{S}^{\dagger} \hatn{S} = \mathbb{I}$ and maps incoming states onto outgoing states, providing all the information of how the full system changes in a collision. We add that one can prove mathematically, under mild conditions, that any state in the Hilbert space $\cal H$ can be interpreted as a free (incoming or outgoing) state. In other words, the domain of the M{\o}ller and the scattering operator is the whole Hilbert space $\cal H$ \cite{Taylor2006}. However, one is usually interested in incoming states where the incident particle is localized in space, far from the collision region and is approaching the scatterer in a state either from  the left or from the right. These are the incoming states considered in this paper.

In our approach, the incoming and outgoing states are density operators associated to the Hilbert space $\cal{H}$. Namely, if the incoming state is factorized $\rho=\hatn{\rho}_X \otimes \hatn{\rho}_{Y}$, then the outgoing state of the full system is $\rho'=S (\hatn{\rho}_X \otimes \hatn{\rho}_{Y}) S^{\dagger}$, and the effect of the collision on the internal state of the system is given by
	\begin{equation}
	\label{stateYgeneral}
	\hatn{\rho}_{Y}' = \mathrm{Tr}_{X} \Big[ \hatn{S} \big( \hatn{\rho}_X \otimes \hatn{\rho}_{Y} \big) \hatn{S}^{\dagger} \Big] \equiv \mathbb{S}\, \hatn{\rho}_{Y} \; ,
	\end{equation}
where $\mathrm{Tr}_X$ denotes the partial trace over $X$. Finally, the system $Y$ can sequentially collide with a stream of particles in identical states $\rho_X$.  Between collisions, the system $Y$ evolves isolated. The dynamics is then obtained by the concatenation
\begin{equation}
\label{concatenation}
\hatn{\rho}_Y^{(n)}=\mathcal{E}_{\tau_n}\circ\mathbb{S}\circ\cdots\circ\mathcal{E}_{\tau_2}\circ\mathbb{S}\circ\mathcal{E}_{\tau_1}\circ\mathbb{S}\,\hatn{\rho}_Y^{(0)},
\end{equation}
where $\tau_i$ is the time between collision $i$ and $i+1$ and $\mathcal{E}_t(\cdot)=e^{-\frac{it}{\hbar}H_Y}(\cdot)e^{\frac{it}{\hbar}H_Y}$ is the unitary map associated to the free evolution with Hamiltonian $H_{Y}$.
 
In the rest of the paper, we calculate and explore the main properties of the superoperator 
$\mathbb{S}$ defining the dynamics in Eqs. \eqref{stateYgeneral} and \eqref{concatenation}.
	
\section{The map in terms of the scattering matrix} \label{sec.ScatMap}

The scattering operator commutes with the free Hamiltonian $[S,H_{0}]=0$ (see appendix \ref{sec:appendixintertwining}), implying that the energy is conserved in a collision. As a consequence, it is convenient to express $S$ in terms of the eigenstates of $H_{0}$. We denote the eigenstates of $H_{Y}$ by $\ket{j}\in {\cal H}_{Y}$:
\begin{equation}
H_{Y}\ket{j}=e_{j}\ket{j} \; ,
\end{equation}
where the eigenvalue $e_{j}$ is the internal energy of the system, and consider the eigenvalues of $H_{Y}$ ordered as $e_1\leq e_2\leq \dots \leq e_N$, denoting the Bohr frequencies by $\Delta_{jk}/\hbar$ with $\Delta_{jk}\equiv e_j-e_k$. Then the generalized eigenstates of $H_{0}$ are the tensor products $\ket{p,j}\equiv\ket{p}\otimes \ket{j}$, where $\ket{p}$ is a plane-wave, i.e., an improper (non-normalizable) state of ${\cal H}_{X}$, whose position representation reads
\begin{equation}\label{planewavex}
\braket{x|p} = \frac{e^{i p x / \hbar}}{ \sqrt{2 \pi \hbar}} \; ,
\end{equation}
and satisfies the generalized orthogonality condition $\braket{p'|p}=\delta(p'-p)$.
The eigenvalue equation for $\hatn{H_0}$ reads
\begin{align}
\hatn{H}_0 \ket{p,j}  & = \left(\frac{\hatn{p}^2}{2m} \otimes \hatn{\mathbb{I}}_Y + \hatn{\mathbb{I}}_X \otimes {H}_Y \right)\ket{p,j} \nonumber \\ & = (E_p + e_j) \ket{p,j} \; ,
\end{align} 
$E_p \equiv p^2/2m$ being the kinetic energy of the plane-wave $\ket{p}$.

Due to the conservation of energy, the elements of the scattering operator $S$ in the eigenbasis of $H_{0}$, $\bra{p',j'}\hatn{S}\ket{p,j}$, are proportional to $\delta(E_{p}-E_{p'}-\Delta_{j'j})$ (see appendix \ref{sec:appendixintertwining}). We express these elements as
\begin{equation}
	\label{expresionS}
	\bra{p',j'}\hatn{S}\ket{p,j}=\frac{\sqrt{|pp'|}}{m}\delta(E_{p}-E_{p'}-\Delta_{j'j}) s_{j'j}^{(\alpha' \alpha)}(E_{p}+e_{j})
\end{equation}
where $s_{j'j}^{(\alpha' \alpha)}(E)$ is an element of the so-called scattering matrix with $\alpha={\rm sign}(p)$ and $\alpha'={\rm sign}(p')$ accounting for the initial and final direction of the momenta, which can be positive ($\alpha,\alpha'=+$) or negative ($\alpha,\alpha'=-$). These elements can be calculated by solving the stationary Schr\"odinger equation with appropriate asymptotic boundary conditions (see appendices \ref{sec:appendixtoperator} and \ref{sec:appendixscattstates}). Importantly, Eq. \eqref{expresionS} defines $s_{j'j}^{(\alpha' \alpha)}(E)$ only for $E\geq {\rm max}\{e_j,e_{j'}\}$ because $p^2$ and $p'^2$ are non-negative. If $j$ and $j'$ fulfill this condition for a given value of $E$, we say that the transition $\ket{j}\to\ket{j'}$ is an open channel. Consequently, the dimension of the matrices ${\bf s}^{(\alpha' \alpha)}(E)$ is $N_{\rm open}(E)\times N_{\rm open}(E)$, where $N_{\rm open}(E)$ is the number of open channels for an energy $E$. For example, if $e_1<E<e_2$, there is only one open scattering channel ($j=j'=1$) and, if $E>e_N$, all channels are open.

The scattering matrix is then ordered in four blocks 
\begin{equation}
\mathbf{s}(E)=\left(
\begin{array}{cc}
\hat{\mathbf{r}}^L(E) & \hat{\mathbf{t}}^R(E) \\
\hat{\mathbf{t}}^L(E) & \hat{\mathbf{r}}^R(E)
\end{array}\right),
\label{matrizs}
\end{equation}
all of dimension $N_{\rm open}(E)\times N_{\rm open}(E)$.
The entries of the transmission-from-the-left matrix $\hat{\mathbf{t}}^L(E)$ are $s_{j'j}^{(+ +)}(E)$, those of the transmission-from-the-right matrix $\hat{\mathbf{t}}^R(E)$ are $s_{j'j}^{(- -)}(E)$, those of the reflection-from-the-left matrix $ \hat{\mathbf{r}}^L(E)$ are
$s_{j'j}^{(- +)}(E)$ , and those of the reflection-from-the-right matrix $ \hat{\mathbf{r}}^R(E)$ are
$s_{j'j}^{(+ -)}(E)$ .
They define the conditional probabilities 
	\begin{align}
	 P^L_{j'j}(E) \equiv  |\hat{t}^L_{j'j}(E)|^2+|\hat{r}^L_{j'j}(E)|^2 
	 &=\sum_{\alpha=\pm} \Big| s_{j'j}^{(\alpha\, +)}(E) \Big|^2  
	 \label{scattmatrixprob}
	\end{align}
for a transition $\ket{j}\to\ket{j'}$ in the system $Y$ given that the momentum of the plane-wave is positive (coming from the left) and the total energy is $E$. We similarly define $P^R_{j'j}(E)$ for a negative momentum (coming from the right). The normalization $\sum_{j'} P_{j'j}^L=\sum_{j'} P_{j'j}^R=1$ follows from the unitary property of $\mathbf{s}(E)$, which corresponds to the diagonal elements of the relation $\mathbf{s}^\dag(E)\mathbf{s}(E)=\mathbb{I}$ satisfied by the scattering matrix Eq.~(\ref{matrizs}), as shown in appendix \ref{sec:appendixsmatrix}.

If we express the density matrix of the system in the eigenbasis of $H_{Y}$, i.e. $(\rho_Y)_{jk}\equiv\braket{j|\rho_{Y}|k}$ and $(\rho'_Y)_{jk}\equiv\braket{j|\rho'_{Y}|k}$, we write Eq.~\eqref{stateYgeneral} as
	\begin{equation}
	\label{stateY}
	(\rho_Y')_{j'k'}= \sum_{j,k}\mathbb{S}^{jk}_{j'k'} (\rho_Y)_{jk} 
	\end{equation} 
with the scattering map
\begin{equation}
\mathbb{S}^{jk}_{j'k'}=\bra{j'}\mathrm{Tr}_{X} \Big[ \hatn{S} \big( \hatn{\rho}_{X} \otimes \ket{j}\bra{k} \big) \hatn{S}^{\dagger} \Big]\ket{k'}.
\label{map-appx}
\end{equation}

	
	\begin{widetext}
	
The superoperator $\mathbb{S}$	can be expressed in terms of the scattering matrix by introducing in Eq.~(\ref{map-appx}) the decomposition 
	$\hatn{\rho}_{X}=\int dp\, dp''\rho_X(p,p'')\ket{p}\bra{p''}$ in the eigenbasis of $H_X$:				
	\begin{equation}
	\label{int-rep}
	\mathbb{S}^{jk}_{j'k'}=\int_{-\infty}^{\infty}  dp'\int_{-\infty}^{\infty}  dp\int_{-\infty}^{\infty}  dp''\rho_X(p,p'')  \bra{p', j'}\hatn{S}\ket{p, j}\bra{p'',k}\hatn{S}^\dagger\ket{p', k'} \; .
	\end{equation}
	 Substituting Eq.~(\ref{expresionS}) in Eq.~(\ref{int-rep}) and using $dE_{p'}=|p'| dp'/m$ we obtain	  
	\begin{equation}
	\mathbb{S}^{jk}_{j'k'}=\sum_{\alpha'=\pm}\int dE_{p'}\,dp \,dp''\rho_X(p,p'')\,\frac{\sqrt{|pp''|}}{m}
	\delta(E_{p}-E_{p'}-\Delta_{j'j})s_{j'j}^{(\alpha'\alpha)}(E_p+e_j)
	\delta(E_{p''}-E_{p'}-\Delta_{k'k})\left[s_{k'k}^{(\alpha'\alpha'')}(E_{p''}+e_{k})\right]^*
	\end{equation}
where the integrals over the momenta $p,p''$ run along the entire real axis, whereas the integral over the energy $E_{p'}$ runs from 0 to infinity and the sum over $\alpha'$ accounts for the positive and negatives values of $p'$. We note that $s_{j'j}^{(\alpha'\alpha)}$ and $s_{k'k}^{(\alpha'\alpha'')}$ are well defined only if the channels $\ket{j}\to \ket{j'}$ and $\ket{k}\to \ket{k'}$ are open at the energies in their arguments, i.e. if $E_p+e_j\geq {\rm max}\{e_j,e_{j'}\}$ and $E_{p''}+e_k\geq {\rm max}\{e_k,e_{k'}\}$. These conditions are equivalent to $E_{p}\geq \Delta_{j'j}$ and $E_{p''}\geq \Delta_{k'k}$, respectively, and are enforced by the delta functions. Integration over the energy $E_{p'}$ yields
	\begin{equation}
	\label{scattmapY}
	\mathbb{S}_{j'k'}^{jk} = \sum_{\alpha'=\pm}\int dp \, dp'' \rho_{X}( p, p'') \,\frac{\sqrt{|pp''|}}{m}\delta(E_{p}-E_{p''}-\Delta_{j'j} + \Delta_{k'k})  s^{(\alpha'\alpha)}_{j'j}(E_{p} + e_j) \left[s^{(\alpha'\alpha'')}_{k'k}(E_{p''} + e_k)\right]^*  .
	\end{equation}
where the integrals run over the region where  $E_{p}\geq\Delta_{j'j}$ and $E_{p''}\geq\Delta_{k'k}$.

\end{widetext}

\begin{widetext}

\section{wave-packets and decoherence} \label{sec.wp}

In this section we study the properties of the scattering map \eqref{stateY} when the incoming particle is in a pure state, i.e., $\hatn{\rho}_X = \ket{\phi}\bra{\phi}$. We will use the momentum representation $\phi(p)\equiv  \braket{p|\phi}$ with $\int dp |\phi(p)|^{2}=1$. To simplify, we consider that all components of $\ket{\phi}$ travel to the right, i.e., $\phi(p)=0$ if $p<0$. For this pure state, $\rho_{X}(p,p'')=\phi(p)\phi^*(p'')$ and Eq.~\eqref{scattmapY} reads
\begin{equation}
\label{intpp'}
\mathbb{S}_{j'k'}^{jk} =  \sum_{\alpha'=\pm} \int dp \, dp'' \phi(p)\phi^*(p'')\frac{\sqrt{pp''}}{m} \delta(E_p-E_{p''} -\Delta_{j'j} +\Delta_{k'k})s^{(\alpha' +)}_{j'j}(E_{p} + e_j)\left[s^{(\alpha' +)}_{k'k}(E_{p''} + e_k)\right]^* \; .
\end{equation}
Recall that  the integration domain in this expression is defined by the inequalities  $E_{p}\geq\Delta_{j'j}$ and $E_{p''}\geq\Delta_{k'k}$.
Since $dE_{p''}=|p''|dp''/m$, the integration of the delta function over $p''$ yields
\begin{equation}
\label{SY-coh}
\mathbb{S}_{j'k'}^{jk} =\sum_{\alpha'=\pm}\int_{ p_{\rm inf}}^\infty dp\, \phi(p)\phi^*\left(\pi(p)\right)\,\sqrt{\frac{p}{\pi(p)}}\,s_{j'j}^{(\alpha' +)}(E_p+e_j)\left[s_{k'k}^{(\alpha' +)}(E_{p}-\Delta_{j'j}+e_{k'})\right]^* 
\end{equation}
\end{widetext}
with  
$\pi(p)=\sqrt{p^{2}-2m(\Delta_{j'j}-\Delta_{k'k})}$.
The  lower integration limit 
$ p_{\rm inf}$ is obtained from $ p^2_{\rm inf}/2m={\rm max}\{0,\Delta_{j'j},\Delta_{j'j}-\Delta_{k'k}\}$, which guarantees that the channels are open in the integration domain.

We now focus on wave-packets centered at a momentum $p_{0}$ and with a width $2\Delta p$, that is, states where the function $\phi(p)$ is zero except for $p\in [p_{0}-\Delta p,p_{0}+\Delta p]$. The properties of the map \eqref{SY-coh} depend crucially on the product
\begin{equation}
\label{prod2}
\phi(p)\phi^*(\pi(p)) \; ,
\end{equation}
which is different from zero if and only if the arguments are within the support of $\phi$, that is,
\begin{equation}
\begin{gathered}
p_{0}-\Delta p < p<p_{0}+\Delta p \\
p_{0}-\Delta p < \pi(p)<p_{0}+\Delta p \; .
\end{gathered}
\end{equation}
Squaring the two inequalities, we have
\begin{equation}
\label{ineq2}
\begin{gathered}
(p_{0}-\Delta p)^{2} < p^{2}<(p_{0}+\Delta p)^{2} \\
(p_{0}-\Delta p)^{2} < p^{2}-2m(\Delta_{j'j}-\Delta_{k'k})<(p_{0}+\Delta p)^{2}
\end{gathered}
\end{equation}
and, eliminating $p$, one gets the following necessary condition for the product \eqref{prod2} to be different from zero for some value of $p$:
\begin{align}
\label{narrow_cond}
 (p_{0}+\Delta p)^{2}-(p_{0}-\Delta p)^{2}&>2m|\Delta_{j'j}-\Delta_{k'k}| \nonumber \\
\Rightarrow \quad \Delta p &>\frac{m|\Delta_{j'j}-\Delta_{k'k}|}{2p_{0}} \; . 
\end{align}
This inequality defines an important distinction between two types of incoming wave-packets: {\it 
a)} Those where the function $\phi(p)$ is highly peaked around $p_{0}$ so that $\Delta p$  
verifies \eqref{narrow_cond} only when $\Delta_{j'j}= \Delta_{k'k}$. Consequently, in \eqref{SY-coh} the term $\mathbb{S}_{j'k'}^{jk}$ vanishes except for transitions with equal energy change. From now on, we call them {\it 
narrow wave-packets}. {\it b)} Those where the function $\phi(p)$ is broad enough to verify 
\eqref{narrow_cond} for at least a pair of transitions $\ket{j}\to \ket{j'}$ and $\ket{k}
\to\ket{k'}$ with $\Delta_{j'j}\neq \Delta_{k'k}$. We call them {\it broad wave-packets} and they 
allow for an overlap of the two factors in \eqref{prod2} for some value of $p$; hence the corresponding term $\mathbb{S}_{j'k'}^{jk}$ in \eqref{SY-coh} can be different from zero.

As we will see below,  narrow (wave-)packets 
destroy most of the coherences or off-diagonal terms of the density matrix of the system, 
whereas they survive or are even created after repeated collisions with broad packets.

This is one of the main results of the paper. It is worth giving a physical interpretation of condition \eqref{narrow_cond} and the distinction between narrow and broad wave-packets. In order to do that, suppose that the initial state of the global system is the pure state $\ket{\phi}\otimes\ket{j}$. After the collision, the  state 
can be written as
	\begin{align}
S\ket{\phi}\otimes\ket{j}&=\sum_{j'}\int dp' \ket{p', j'}\bra{p', j'}\,S\,\left[\ket{\phi}\otimes\ket{j}\right] \nonumber \\
&=\sum_{j'}\int dp'dp \ket{p' ,j'} \bra{p', j'}S\ket{p,j}\phi(p) \; .
	\end{align}
Using  Eq.~\eqref{expresionS} and integrating the delta function, the outgoing state can be written as
\begin{equation}
S\ket{\phi}\otimes\ket{j}=\sum_{j'}\sum_{\alpha'=\pm} \ket{ \phi_{j'}^{\alpha'}}\otimes\ket{j'} \; . 
\end{equation}
Here $\ket{\phi_{j'}^{\alpha'}}$ are non-normalized wave-packets, whose momentum representation reads
\begin{equation}
\braket{p|\phi_{j'}^{\alpha'}}
=\sqrt{\frac{|p|}{|\bar\pi(p)|}}
s_{j'j}^{(\alpha' +)}(E_{p}+e_{j'})\phi(\bar\pi(p))\Theta (\alpha' p)
\end{equation}
with $\bar\pi(p)\equiv\sqrt{p^2+2m \Delta_{j'j}}$, and $\Theta(p)$ being the Heaviside step function.
Hence, the state after the collision is a superposition of wave-packets $\ket{\phi_{j'}^{\alpha'}}\otimes\ket{j'}$, corresponding to the different transitions $\ket{j}\to \ket{j'}$, that leave the scatterer with positive ($\alpha=+$, transmitted packets) or negative  ($\alpha=-$, reflected packets)
momentum. The support of this outgoing packets is given by the function $\phi$ and is determined by the inequalities
\begin{equation}
p_0-\Delta p<\sqrt{p^2+2m \Delta_{j'j}}<p_0+\Delta p
\end{equation}
with $p>0$ for the transmitted packets and $p<0$ for the reflected ones.

\begin{figure}[h]
	\centering
	\includegraphics[scale=0.35]{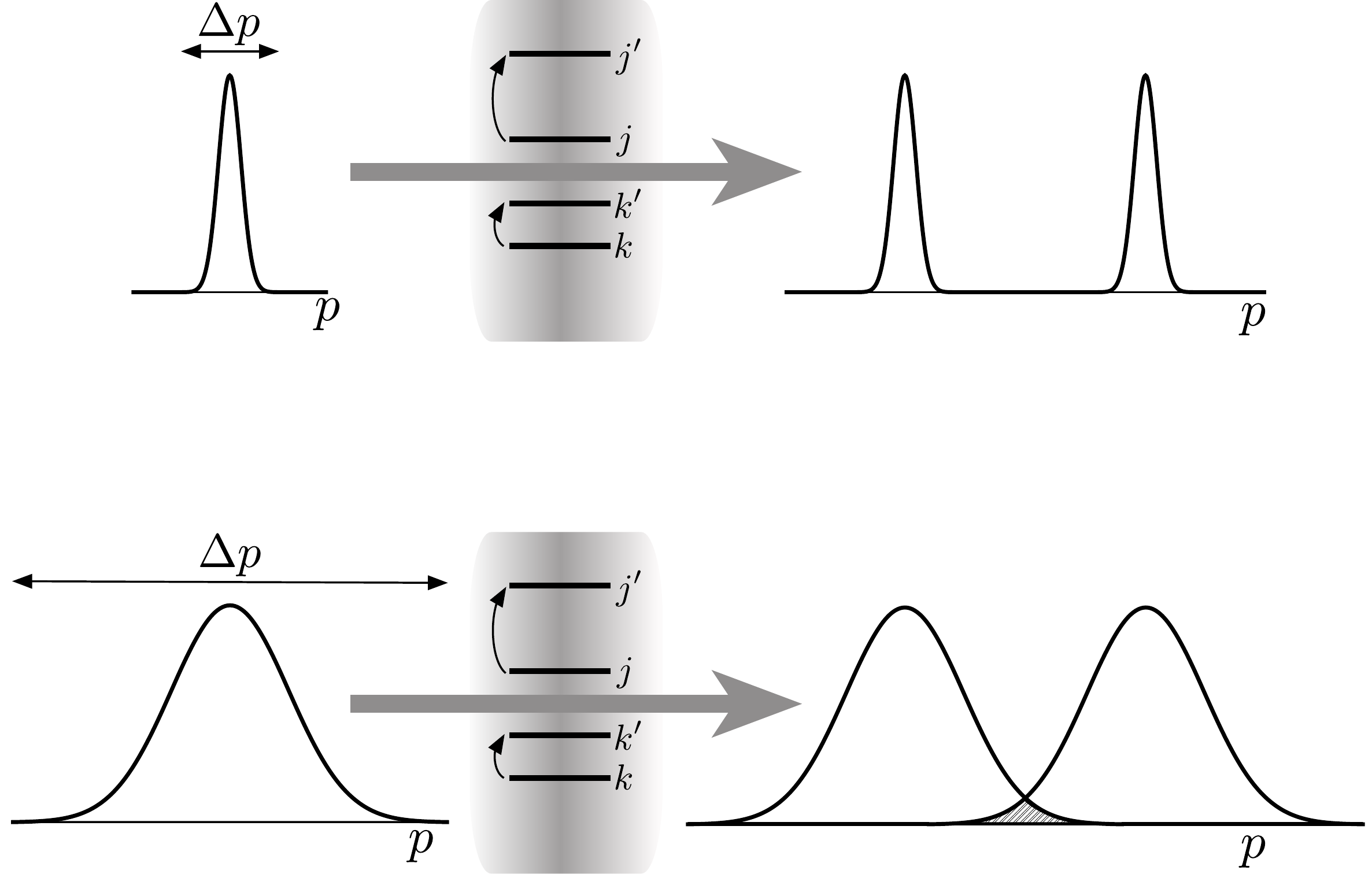}
	\caption{Narrow (upper plate) and broad  (lower plate) wave-packets. The blurred grey zones represent the scatterer, which undergoes the transitions $\ket{j}\to\ket{j'}$ and $\ket{k}\to\ket{k'}$. The figure shows the two transmitted wave-packets corresponding to these specific transitions (notice that the collision will generate, in general, much more outgoing packets than the two depicted). The crucial difference is that the overlap of the outgoing wave-packets is zero in the narrow case, unless both have exactly the same energy: $\Delta_{j'j}=\Delta_{k'k}$. For broad packets, coherences between states do not vanish and can even be created due to the overlap (dark area) of the two outgoing packets depicted in the lower plate. \label{narrow_broad}}
	\end{figure}

Consider now the initial states $\ket{\phi}\otimes\ket{j}$ and $\ket{\phi}\otimes\ket{k}$ and the corresponding transitions, $\ket{j}\to\ket{j'}$ and $\ket{k}\to\ket{k'}$. The transmitted wave-packets overlap in the momentum representation if there is a positive $p$ such that 
\begin{equation}\label{narrow_cond2}
\begin{split}
p_0-\Delta p&<\sqrt{p^2+2m \Delta_{j'j}}<p_0+\Delta p \\
p_0-\Delta p&<\sqrt{p^2+2m \Delta_{k'k}}<p_0+\Delta p.
\end{split}
\end{equation}
It is straightforward to prove that these two conditions are equivalent to \eqref{narrow_cond}. 
However, Eq.~\eqref{narrow_cond2}  provides an illuminating interpretation of the distinction between narrow and broad wave-packets, which is  sketched in Fig.~\ref{narrow_broad}. In the case of an incident narrow packet, 
condition \eqref{narrow_cond2} means that the outgoing wave-packets resulting from channels 
$\ket{j}\to \ket{j'}$ and $\ket{k}\to \ket{k'}$ do not overlap unless they induce jumps in the 
system state with exactly the same energy, that is, $\Delta_{j'j}=\Delta_{k'k}$; in such a case, 
the outgoing packets are identical. On the other hand, for a broad packet obeying \eqref{narrow_cond2} for a pair of transitions with $\Delta_{j'j}\neq\Delta_{k'k}$, outgoing wave-packets 
with different energy may overlap, as shown in the lower diagram of Fig.~\ref{narrow_broad}. If 
this occurs, then the collision preserves the coherences between states $\ket{j}$ and $\ket{k}$ 
and can even create new coherences when a single pure state $\ket{j}$ jumps to a superposition 
of $\ket{j'}$ and $\ket{k'}$ and the resulting wave-packets overlap, as we show below in detail.

The same results apply to a wave-packet where $\phi(p)$ decays and is almost zero far from its center at $p_{0}$. We will consider, for instance,  Gaussian wave-packets $\ket{\phi_{p_0,x_0}}$ of the form:
	\begin{equation}
	\label{gaussiandistribution}
 \braket{p|\phi_{p_0,x_0}} = (2 \pi \sigma^2)^{-1/4} \exp \Big[-\frac{(p-p_0)^2}{4\sigma^2} - i\, \frac{px_0}{\hbar} \Big],
	\end{equation}
where $p_0=\bra{\phi_{p_0,x_0}}\hatn{p}\ket{\phi_{p_0,x_0}}$ is the expectation value of the momentum,  $\sigma>0$  its standard deviation, and $x_0=\bra{\phi_{p_0,x_0}}\hatn{x}\ket{\phi_{p_0,x_0}}$ is the expectation value of the position operator. The condition $ \phi_{p_0,x_0}(p)=0$ for $p<0$, which we have assumed above, is satisfied if $p_{0}$ is positive and  $p_0\gg\sigma$. The previous discussion applies to Gaussian wave-packets, for which the condition \eqref{narrow_cond} for a narrow packet is
\begin{equation}\label{narrow_cond_gauss}
\sigma \ll \frac{m|\Delta_{j'j}-\Delta_{k'k}|}{2p_{0}}
\end{equation}
for every quadruplet $\{j,k,j',k'\}$ with $\Delta_{k'k}\neq \Delta_{j'j}$.

\subsection{Narrow wave-packets}

%

As discussed previously, a consequence of Eq.~\eqref{narrow_cond} is that, in a collision with an incident narrow packet, elements $(\rho_Y)_{jk}$ contribute to $(\rho'_Y)_{j'k'}$ only if $\Delta_{j'j}=\Delta_{k'k}$. In this case, $\pi(p)=p$ in Eq.~\eqref{SY-coh}. Moreover, since the packet is narrow, we can assume that the scattering matrix is approximately constant in the support of $\phi(p)$ and that $|\phi(p)|^{2}$ is approximately normalized in the integration domain $[p_{\rm inf},\infty)$. Under these assumptions, Eq.~\eqref{SY-coh} reduces to
\begin{align}
\mathbb{S}_{j'k'}^{jk} &\simeq\sum_{\alpha'=\pm}s_{j'j}^{(\alpha' +)}(E_{p_{0}}+e_j)\left[s_{k'k}^{(\alpha' +)}(E_{p_{0}}+e_{k})\right]^*  \nonumber \\
 &=
\hat t^L_{j'j}(E_{p_0}+e_j)\left[\hat t^{L}_{k'k}(E_{p_0}+e_k)\right]^* \nonumber 
\\ &
+\hat r^L_{j'j}(E_{p_0}+e_j)\left[ \hat r^{L}_{k'k}(E_{p_0}+e_k)\right]^* ,\label{SY-Coh2}
\end{align}
for $\Delta_{j'j}=\Delta_{k'k}$ and zero otherwise. Let us apply this condition first to the diagonal terms of the density matrix $(\rho_{Y})_{jj}$, which are the populations of the energy levels. The condition  $\Delta_{j'j}=\Delta_{k'k}$  for $j'=k'$ implies $e_{j}=e_{k}$. If $H_{Y}$ is non-degenerate, then $j=k$ and
\begin{align}
\label{SY-pop}
\mathbb{S}_{j'j'}^{jj} & = \left[|\hat t^{L}_{j'j}(E_{p_0} + e_j)|^2+|\hat r^{L}_{j'j}(E_{p_0} + e_j)|^2\right] \nonumber \\ & = P^{L}_{j'j}(E_{p_0} + e_j)
\end{align}
where the transition probability $P^{L}_{j'j}(E)$ is the one defined in Eq.~\eqref{scattmatrixprob}.
We see that  the evolution of the populations or diagonal terms of the density matrix $(\rho_{Y})_{jj}$ is independent of the off-diagonal terms or coherences. The diagonal terms obey the master equation:
\begin{equation}
(\rho'_{Y})_{j'j'}=\sum_{j} P^{L}_{j'j}(E_{p_0} + e_{j}) (\rho_{Y})_{jj}
\label{eq.32}
\end{equation}
which conserves the trace or total probability since $\sum_{j'}P^{L}_{j'j}(E)=1$, due to the unitarity of the scattering matrix ${\bf s}(E)$ [see the discussion below Eq.~\eqref{scattmatrixprob}].

Let us discuss the evolution of the coherences or off-diagonal terms of the density matrix 
$\rho_{Y}$, and suppose first that $H_Y$ is non-degenerate and the Bohr frequencies of the system are non-degenerate,
i.e., the only solutions to $\Delta_{j'j}=\Delta_{k'k}$ are $j'=j$ and $k'=k$  or $j'=k'$ and 
$j=k$. The second case corresponds to the evolution of the populations that we have already
discussed. The first case corresponds to $\mathbb{S}^{jk}_{jk}$ and yields the following evolution equation for the off-diagonal terms:
\begin{equation}\label{cohere43}
(\rho'_Y)_{jk}=\mathbb{S}^{jk}_{jk}(\rho_Y)_{jk}\qquad \mbox{for $j\neq k$.}
\end{equation}
Therefore, each coherence evolves independently and is simply  changed by a multiplicative factor after a collision. Moreover, since $|t^{L}_{j'j}(E)|^2+|r^{L}_{j'j}(E)|^2\leq 1$ for any $E$, $j$, and $j'$, a direct application of the  Cauchy-Schwarz inequality yields
\begin{align}
&\left|t^L_{j'j}(E)t^{L}_{k'k}(E')^* +r^L_{j'j}(E)r^{L}_{k'k}(E')^* \right|\nonumber \\
&\leq \left(|t^{L}_{j'j}(E)|^2+|r^{L}_{j'j}(E)|^2\right)\left(|t^{L}_{k'k}(E')|^2+|r^{L}_{k'k}(E')|^2\right)\nonumber \\
&\leq 1
\end{align}
for all $j,j',k,k'$, $E$ and $E'$. We conclude that all the entries of the scattering map are
bound as $|\mathbb{S}^{jk}_{j'k'}|\leq 1$, and, according to Eq.~\eqref{cohere43}, coherences
either decay or remain finite only if $|\mathbb{S}^{jk}_{jk}|= 1$. The latter case happens
only  when the two equalities $|t^{L}_{aa}(E)|^2+|r^{L}_{aa}(E)|^2=1$ for $a=j,E=E_{p_0}+e_j$
and $a=k,E=E_{p_0}+e_k$ are simultaneously satisfied. This is extremely unlikely in a 
multi-channel scattering process. Therefore, generically one observes the strict inequality, which leads to decoherence in systems without degenerate Bohr frequencies.

%
%

Summarizing, we found that decoherence is generic for systems  whose Bohr frequencies are non-degenerate. 
Coherences among degenerate Bohr frequencies evolve uncoupled to the other elements of the density matrix but do not necessarily decay. 
Moreover, the only possible coupling between coherences and populations arises in systems with degenerate energy spectrum. In this case,  
coherences between degenerate levels (zero Bohr frequency) are mapped to populations and populations are mapped to coherences between degenerate levels.



\subsection{Broad wave-packets}

If the  momentum width of the wave-packet  is large enough, the map $\mathbb{S}$ is more complex and can keep and create coherences. For example, suppose that condition \eqref{narrow_cond} is satisfied for $k=j$ and some $j'\neq k'$, i.e.
\begin{equation}\label{narrow_cond3} 
\Delta p >\frac{m|\Delta_{j'j}-\Delta_{k'j}|}{2p_{0}}.
\end{equation}
In this case, the term $\mathbb{S}^{jj}_{j'k'}$ is different from zero. If the system is initially in the pure state $\rho_{Y}=\ket{j}\bra{j}$ then, after the collision, we have
\begin{equation}
(\rho'_{Y})_{j'k'}=\mathbb{S}^{jj}_{j'k'}\neq 0,
\end{equation}
that is, a non zero coherent or off-diagonal term. 

%

Therefore, broad wave-packets can induce coherences in a system initially diagonal in the energy basis, while narrow wave-packets generically lead to decoherence  in non-degenerate systems. In Sec.~\ref{sec.appli}, we explore this crucial difference in specific examples.

\section{Ensembles of wave-packets and thermalization} \label{sec.ensemble}

In this section we consider that the initial state of $X$ is given by a statistical ensemble of wave-packets and discuss sufficient conditions that lead to thermalization for system $Y$. Here, we assume that $[H_Y,\nu]\neq 0$ to ensure that $[S,\hatn{\mathbb{I}}_X \otimes \nu] \neq 0$, otherwise the scattering process induces transitions only between the (common) eigenstates of $H_Y$ and $\nu$ with different eigenvalues, ruling out the possibility of thermalization of $Y$ by repeated collisions.

The first condition for thermalization is that the ensemble consists of narrow wave-packets. As we have just seen, narrow wave-packets lead to decoherence while broad ones induce coherences even if the initial state of the system is diagonal in the energy basis. This remains true for an ensemble of packets described by a density matrix,  as we illustrate in the examples of Sec.~\ref{sec.appli}.

\subsection{Micro-reversibility}

The second condition is microscopic reversibility, that is, the invariance of the scattering operator under time reversal. In quantum mechanics, time reversal is implemented by an anti-unitary operator $\textsf{T}$ which changes the sign of all momenta and other odd magnitudes under time reversal, like angular momentum, spin, and the magnetic field. It takes different forms, depending on the system. For instance, for a spinless point particle, $\textsf{T}$ is the conjugation of the wave function in position representation: $\braket{x|\textsf{T}|\psi}=\braket{x|\psi}^{*}$. In the momentum representation, on the other hand, the operator changes the sign of the momenta: $\braket{p|\textsf{T}|\psi}=\braket{-p|\psi}$.
If the free and the total Hamiltonian commute with $\textsf{T} = \textsf{T}_X \otimes \textsf{T}_Y$, where $\textsf{T}_X$ and $\textsf{T}_Y$ are the time reversal operators acting on the Hilbert space of $X$ and $Y$, then the scattering operator also commutes with $\textsf{T}$ and the collision is invariant under time reversal. For simplicity, we assume here that the eigenstates of $H_{Y}$ are also invariant, that is, 
$\textsf{T}_Y\ket{j}=\ket{j}$. In this case, the scattering matrix obeys the following symmetry relation:
\begin{equation}\label{micro-rev2}
s_{j'j}^{(\alpha'\, \alpha)}(E)=s_{jj'}^{(-\alpha\, -\alpha')}(E)
\end{equation}
or equivalently  $\hat{\mathbf{r}}^L=(\hat{\mathbf{r}}^{L})^t$, $\hat{\mathbf{r}}^R=(\hat{\mathbf{r}}^{R})^t$ and  $\hat{\mathbf{t}}^L=(\hat{\mathbf{t}}^{R})^t$ where $(\cdot)^t$ stands for transpose \cite{Belkic2004,Taylor2006,Nazarov2009}. In other words, (\ref{matrizs}) is a symmetric matrix. 

 \begin{figure}[t]
	\[
	\includegraphics[scale=0.58]{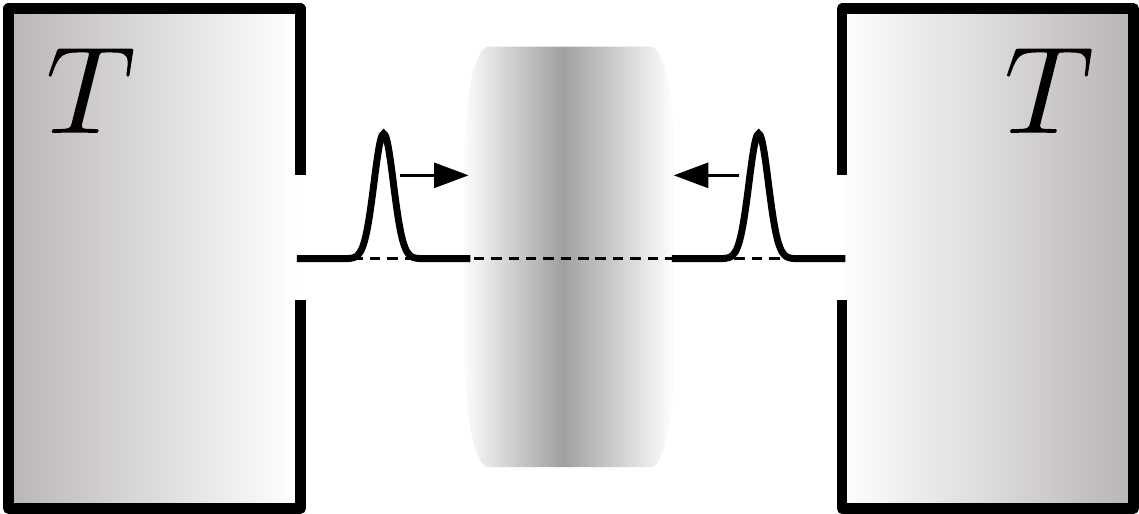}
	\]
\caption{Thermalization via collisions is achieved if the system (scatterer) is bombarded by narrow wave-packets coming from equilibrium reservoirs at the two sides. If we remove one of the reservoirs, the setup is out of equilibrium. Notice however that, if the scatterer is symmetric, that is, if the interaction potential $V(x)$ is even, $V(x)=V(-x)$, then the second reservoir is not necessary. \label{figthermalrev}}
	\end{figure}

In our previous discussion, we only  consider particles incident from the left. This setting is not time-reversal invariant. To satisfy microscopic reversibility, we have to consider particles coming  from both left and right, as shown schematically in Fig.~\ref{figthermalrev}. 
That is, for every wave-packet 
$\ket{\phi_{p_0,x_0}}$ in the ensemble,
centered around the momentum $p_0>0$ and coming from the left ($x_0<0$), there must be a $\ket{\phi_{-p_0,-x_0}}$ centered around momentum $-p_0<0$ and coming from the right ($-x_0>0$). 
For narrow wave-packets, the position $x_0$ becomes irrelevant and we omit it from the notation. 
With these ideas in mind, we consider a symmetric ensemble defined by the  probability distribution $\mu(p_{0})$, normalized in $[0,\infty)$:
\begin{equation}
\hatn{\rho}_X=\int_{0}^{\infty} dp_0 \frac{\mu(p_0)}{2}(\ket{\phi_{p_0}}\bra{\phi_{p_0}}+\ket{\phi_{-p_0}}\bra{\phi_{-p_0}})
\label{ensemble1}
\end{equation}
Inserting this incoming state into Eq.~(\ref{scattmapY}), and taking into account that all the packets are narrow, one obtains 
\begin{equation} 
\mathbb{S}_{j'j'}^{jj} =
\int dp_0\frac{\mu(p_0)}{2}\left[P^L_{j'j}(E_{p_0} + e_j)+P^R_{j'j}(E_{p_0} + e_j)\right] 
\label{S_Y-TR-aux}
\end{equation}
where the integral over $p_{0}$ runs over all positive values satisfying $E_{p_{0}}\geq \Delta_{j'j}$. Recall that $P^L_{j'j}(E)$ is given by Eq.~(\ref{scattmatrixprob}) and similarly for $P^R_{j'j}(E)$. To simplify the notation, we define 
\begin{equation}
P_{j'j}(E)=\frac{P_{j'j}^L(E)+P_{j'j}^R(E)}{2} 
\label{defP}
\end{equation}
and replace it in Eq.~(\ref{S_Y-TR-aux}), which becomes
\begin{equation}
\mathbb{S}^{jj}_{j'j'}=\int dp_0\,\mu(p_0) P_{j'j}(E_{p_0}+e_j) \; .
\label{S_Y-TR}
\end{equation}
Let us recall that
 the energy in the argument of $P_{j'j}(E)$ is such that the scattering channel $\ket{j}\to\ket{j'}$ is open. Finally, from the definition of the transition probabilities in Eq.~\eqref{scattmatrixprob} and the micro-reversibility condition \eqref{micro-rev2}, we obtain
\begin{align}
\label{TRSymmetry}
P_{j'j}^L(E)+P_{j'j}^R(E)&=P_{jj'}^L(E)+P_{jj'}^R(E)\nonumber \\
\Rightarrow \quad  P_{j'j}(E) &= P_{jj'}(E) \; .
\end{align}

\subsection{Detailed balance and thermalization}

A third condition for the thermalization of the system $Y$ is that the statistics of the narrow wave-packets $X$ is thermal. We show below that, in this case, the map that evolves the populations obeys the detailed balance condition, ensuring thermalization. 

According to the discussion in~\cite{Jannik2019} we take $\mu(p_0)=\mu_{\rm eff}(p_0)$ with
\begin{equation}
\mu_{\rm eff}(p)=\beta\frac{p}{m}e^{-\beta p^2/2m}
\label{effdist.eq}
\end{equation}
the effusion distribution, in which case Eq.~(\ref{S_Y-TR})
satisfies the detailed-balance relation
\begin{equation}
\mathbb{S}^{jj}_{j'j'}e^{-\beta e_j}=\mathbb{S}^{j'j'}_{jj}e^{-\beta e_{j'}} \; .
\label{DB2}
\end{equation}

\begin{widetext}

To prove it, let us compute the left handside using Eq.~\eqref{S_Y-TR}, Eq.~\eqref{SY-coh} and the change of variable $E=E_{p_{0}}+e_{j}$:
\begin{equation}
\mathbb{S}^{jj}_{j'j'} e^{-\beta e_j}=  \int_{p_{{\rm inf}}}^\infty dp_0\frac{\beta\,p_0}{m}
e^{-\beta(E_{p_0}+e_j)} P_{j'j}(E_{p_0}+e_j)
=\beta\int_{{\rm max}\{e_j,e_{j'}\}}^\infty dE\, e^{-\beta E} P_{j'j}(E),
\end{equation}
where $p_{{\rm inf}}$ is $\sqrt{2m\Delta_{j'j}}$ if $\Delta_{j'j} > 0$ and zero otherwise.
Similarly, the right handside of \eqref{DB2} reads
\begin{equation}
\mathbb{S}^{j'j'}_{jj}e^{-\beta e_{j'}}=  \int_{p_{{\rm inf}}}^\infty dp_0\frac{\beta\,p_0}{m}e^{-\beta(E_{p_0}+e_{j'})} P_{jj'}(E_{p_0}+e_{j'})=\beta
\int_{{\rm max}\{e_{j'},e_{j}\}}^\infty dE \,e^{-\beta E} P_{jj'}(E), 
\end{equation}
where now $p_{{\rm inf}}$ is $\sqrt{2m\Delta_{jj'}}$ if $\Delta_{jj'} > 0$ and zero otherwise. 
Both integrands are the same because time-reversal symmetry implies $P_{j'j}(E)=P_{jj'}(E)$. Hence, the detailed-balance equality Eq.~(\ref{DB2}) is satisfied, which, in turn, guarantees the thermalization of system $Y$, that is, after a number of collisions the system reaches the thermal state $\rho_Y = e^{-\beta H_{Y}}/Z_{Y}$ with $Z_Y = \mathrm{Tr}(e^{-\beta H_{Y}})$.

\end{widetext}

Notice that if, on the other hand, we take the Maxwell-Boltzmann distribution $\mu(p_0)=\sqrt{{\beta}/({2m\pi})}\exp[-\beta p_0^2/(2m)]$ in Eq.~(\ref{S_Y-TR}),
$(\mathbb{S}_Y)^{jj}_{j'j'}$ does not satisfy detailed-balance and the map does not thermalize the system $Y$.
The physical reason is that particles escaping from a small hole in a thermal box are distributed in momentum according to the effusion distribution, which is the Maxwell-Boltzmann distribution weighted with a flux factor. See \cite{Jannik2019} for a detailed discussion of this subtle issue.

We end this section with two remarks. First, if the potential $V(x)$ has the spatial symmetry $x\to -x$, then  $\hat{\mathbf{r}}^L =\hat{\mathbf{r}}^R$ and $\hat{\mathbf{t}}^L =\hat{\mathbf{t}}^R$, and
 $P_{j'j}^L(E)=P_{j'j}^R(E)$. The spatial reflection symmetry $x\to -x$ plus time reversal symmetry, Eq.~(\ref{TRSymmetry}), imply $P_{j'j}^L(E)=P_{jj'}^L(E)$. In this case, detailed balance (and therefore thermalization) is satisfied with just left (or right) incoming wave-packets.

Second, we have seen that the narrow packets destroy coherences in the eigenbasis of $H_{Y}$. Consequently, once the diagonal state is reached, the unitary free evolution between collisions present in Eq.~\eqref{concatenation} does not change the density matrix $\rho_{Y}$. We conclude that narrow wave-packets induce thermalization independently of the time intervals between collisions, which can be either random or deterministic.

\subsection{Entropy production}

The evolution of populations  $p_j\equiv (\rho_Y)_{jj}$  in  system $Y$ is ruled by the discrete-time stochastic master equation 
\begin{equation}
p'_{j}=\sum_k \mathbb{W}_{jk} p_{k} \; ,
\end{equation}
 where $\mathbb{W}_{jk}\equiv\mathbb{S}^{kk}_{jj}$ satisfies $\sum_j\mathbb{W}_{jk}=1$ and detailed balance  $\mathbb{W}_{jk}e^{-\beta e_k}=\mathbb{W}_{kj}e^{-\beta e_{j}}$. This allows the identification of the average energy change of $Y$ with heat, i.e., $\Delta S\geq \beta \Delta E$ where the entropy $S\equiv -\sum_j p_j\ln p_j$ is given by the Shannon entropy associated to system $Y$, and $E=\sum_j e_j p_j$ its average energy, as shown in~\cite{BarraEsp2016} (see also~\cite{Gaspard04b,Altaner12}). The energy change is determined by
\begin{align}
Q\equiv\Delta E & = \sum_j e_j (p'_j-p_j) \nonumber \\
 & =\sum_{j,k} e_j(\mathbb{W}_{jk} - \delta_{jk})p_k
\label{heat.eq}
\end{align}
that we denote $Q$, anticipating its interpretation as heat. 
We now consider the change in the Shannon entropy of the system $Y$
\begin{equation}
\Delta S=\sum_j (p_j\ln p_j-p'_j\ln p'_j)= \beta Q + \Sigma 
\end{equation}
that we split into two contributions. The first one is the entropy flow given by
\begin{equation}
\beta Q= - \sum_{j,k} \mathbb{W}_{jk} p_{k} \ln \frac{\mathbb{W}_{jk}}{\mathbb{W}_{kj}} \; , 
\end{equation}
where we have used the detailed balance condition $ \ln ({\mathbb{W}_{jk}}/{\mathbb{W}_{kj}} )=-\beta(e_j-e_{k})$ for the identification with Eq.~(\ref{heat.eq}). The second one is 
the entropy production given by
\begin{equation}
\Sigma= \sum_{j,k}  \mathbb{W}_{jk} p_{k} \ln \frac{ \mathbb{W}_{jk} p_{k}}{ \mathbb{W}_{kj} p'_{j}} \; .
\end{equation}
This quantity is non-negative as can be shown using Jensen inequality. Indeed, since  $-\ln x \geq x-1$, 
\begin{align}
\Sigma & \geq \sum_{j,k} \mathbb{W}_{jk} p_{k} 
\left( \frac{\mathbb{W}_{kj} p'_{j}}{\mathbb{W}_{jk} p_k} -1 \right)\nonumber
\\ &=\sum_{j,k} \left( \mathbb{W}_{kj} p'_{j}- \mathbb{W}_{jk} p_{k}  \right) =0 \; .
\end{align}
When the system reaches equilibrium i.e., when $p_j=p'_j=e^{-\beta e_j}/Z_{Y}$, we have $\Sigma=0$. 

\section{Extension to particles with internal structure} \label{sec.comment}

We briefly comment the statement from the introduction that if 
the incoming particle has an internal structure and its state is of the form $\rho_X\otimes \rho_\chi$ with $\rho_X$ associated to the translation degree of freedom and $\rho_\chi$ to the internal structure, the effective system $Y$ comprises $\chi$ and the fixed scatterer, here called $\Upsilon$. The effective system $Y$ before the collision has a state $\rho_Y=\rho_\chi\otimes \rho_\Upsilon$ and the effective Hamiltonian is $H_Y=H_\chi+H_\Upsilon$, with $H_\chi$ the Hamiltonian associated to the internal degrees of freedom of the traveling particle and $H_\Upsilon$ that of the fixed scatterer.  
Narrow wave-packets (with respect to the smallest level spacing of $Y$) induce decoherence in $Y$, and thus also in $\chi$ and $\Upsilon$. 
Thermalization of $\Upsilon$ after many collisions still holds if $\rho_X$ is a thermal ensemble of wave-packets and the internal structure of the particle is also thermal i.e. $\rho_\chi\sim e^{-\beta H_\chi}$. This follows from the detailed balance property of ${\mathbb S}_{j'j'}^{jj}$. The index $j$ stands now for the pair $(\mathsf{x,y})$ where $\mathsf{x}$ indexes the eigenvalues and eigenstates of $H_\chi$ and $\mathsf{y}$ those from $H_\Upsilon$. Due to the product structure of the state of $Y$ prior to the collision, the population of $Y$ evolves with the equation $P_{\mathsf{x'y'}}=\sum_{\mathsf{xy}} {\mathbb S}_{\mathsf{(x'y')(x'y')}}^{\mathsf{(xy)(xy)}} e^{-\beta \epsilon_\mathsf{x}}P_\mathsf{y}$ and the reduced dynamics for the populations of $\Upsilon$ is given by $P_\mathsf{y'}=\sum_\mathsf{y} {\mathbb R}_\mathsf{y'}^{\mathsf{y}}P_\mathsf{y}$ with the reduced stochastic map ${\mathbb R}_\mathsf{y'}^{\mathsf{y}}=\sum_\mathsf{xx'}{\mathbb S}_\mathsf{(x'y')(x'y')}^\mathsf{(xy)(xy)}e^{-\beta \epsilon_\mathsf{x}}$ also satisfying detailed balance and thus inducing thermalization of $\Upsilon$.

 
\section{Applications} \label{sec.appli}

In this section, we illustrate in simple models the results previously obtained. We consider a system $Y$ with finite dimension $N$, i.e., $H_Y=\sum_{j=1}^{N}e_j\ket{j}\bra{j}$ 
interacting with $X$ via the coupling $V(x)\otimes \nu$  in Eq.~(\ref{hamiltonianfullYX}) with $V(x)=g\delta(x)$.  
The scattering problem is solved in appendix \ref{app:ejemplo-delta}. Since the potential is symmetric under the spatial inversion $x\to -x$, the transmission and reflection matrices from the left equal those from the right, i.e. $\hat{\mathbf{r}}^L(E)=\hat{\mathbf{r}}^R(E)\equiv\hat{\mathbf{r}}(E)$ and $\hat{\mathbf{t}}^L(E)=\hat{\mathbf{t}}^R(E)\equiv\hat{\mathbf{t}}(E)$. The resulting scattering matrix in Eq.~(\ref{matrizs}) is given by
\begin{equation}
\hat{t}_{j'j}=\sqrt{\frac{p_{j'}}{p_j}}t_{j'j} \quad{\rm and} \quad \hat{r}_{j'j}=\sqrt{\frac{p_{j'}}{p_j}}(\delta_{j'j}-t_{j'j})
\end{equation}
with $1\leq j,j'\leq N_{\rm open}(E)$, where $N_{\rm open}(E)=\#\{e_{j'}\leq E\}$, the number of levels with energy smaller than $E$, $p_j=\sqrt{2m(E-e_j)}$ and $t_{j'j}$ are the elements of the $N\times N$ matrix
\begin{equation}
\mathbf{t}=[\,\mathbb{I}+\frac{img}{\hbar^2}\,\mathbb{D}^{-1}\mathbb{V}\,]^{-1}.
\label{eq:t-matrix1}
\end{equation}
In Eq.~(\ref{eq:t-matrix1}), the matrix $\mathbb{D}$ is diagonal with  $\mathbb{D}_{jj}=p_j/\hbar$ and $\mathbb{V}$ the matrix with elements $\mathbb{V}_{kj}=\bra{k}\nu\ket{j}.$
We take in the following numerical examples $\hbar = m = g = 1$. 

\subsection{Narrow and broad wave-packets}
We first illustrate the role of the width of the wave-packet, i.e., the transition from narrow to broad wave-packets.
For this, the simplest is to investigate a two-level system $Y$. Let us remind that, according to  Eq.~(\ref{narrow_cond_gauss}), a Gaussian wave-packet is considered to be narrow if $\sigma\ll m \Delta_Y/{2p_0}$ with $\Delta_Y=e_2-e_1.$ We take $\mathbb{V}=\sigma^x+ \,\sigma^z$ 
(the Pauli matrices) in Eq.~(\ref{eq:t-matrix1}) and compute the map $\mathbb{S}$ according to Eq.~(\ref{scattmapY}). Between collisions, we simply take $\mathcal{E}_{\tau_n}$ as the identity for all $n$. 

The results are shown in Fig.~\ref{fig-bloch}, where we plot the state in the Bloch sphere representation according to $\rho_Y = (\mathbb{I} + \vec{P} \cdot \vec{\sigma})/2$ and $\vec{P}$ is the polarization vector. We take two initially pure states (lying on the surface of the sphere), one being in an excited state and the other in a superposition of ground and excited states (Fig.~\ref{fig-bloch}, upper and lower panel, respectively). 

{\it Collisions with narrow wave-packets} lead to decoherence of $Y$ for both initial states, 
see the blue dots converging to the invariant state in the $z$ axis near the center of the 
sphere as the scattering map is iterated. The populations of the invariant state can be computed 
by evaluating the invariant state of Eq.~\eqref{eq.32}. For the two-level system under 
consideration, the scattering matrix ${\mathbf s}(E)$ depends only weakly on $E$ and therefore 
$P^L_{j'j}(E_{p_0}+e_j)\approx P^L_{j'j}(E_{p_0})$ for all $j'j$ in Eq.~\eqref{eq.32}. Since 
$P^L_{j'j}(E)$ is a bi-stochastic matrix, the invariant state of Eq.~\eqref{eq.32} is very close 
to the maximally-mixed state. In Fig.~\ref{fig-bloch}, upper panel, we see (blue dots) that a 
state which is initially diagonal in the energy basis remains diagonal throughout the evolution 
i.e. its dynamics is confined to the $z$-axis. In Fig.~\ref{fig-bloch}, lower panel, an initial 
superposition state with an equal amount of populations dephases and changes its populations in 
an almost negligible way as it spirals towards the invariant state. Conversely, {\it collisions 
with broad wave-packets} transfer coherences to an initially diagonal state (which leaves the 
$z$-axis) and couple the populations and coherences of an initial superposition state (which 
quickly leaves the $xy$-plane). For such broad wave-packets, the system evolves towards a 
steady-state with coherences (red triangles in Fig.~\ref{fig-bloch}).



\begin{figure}[htb]
\includegraphics[width=9cm]{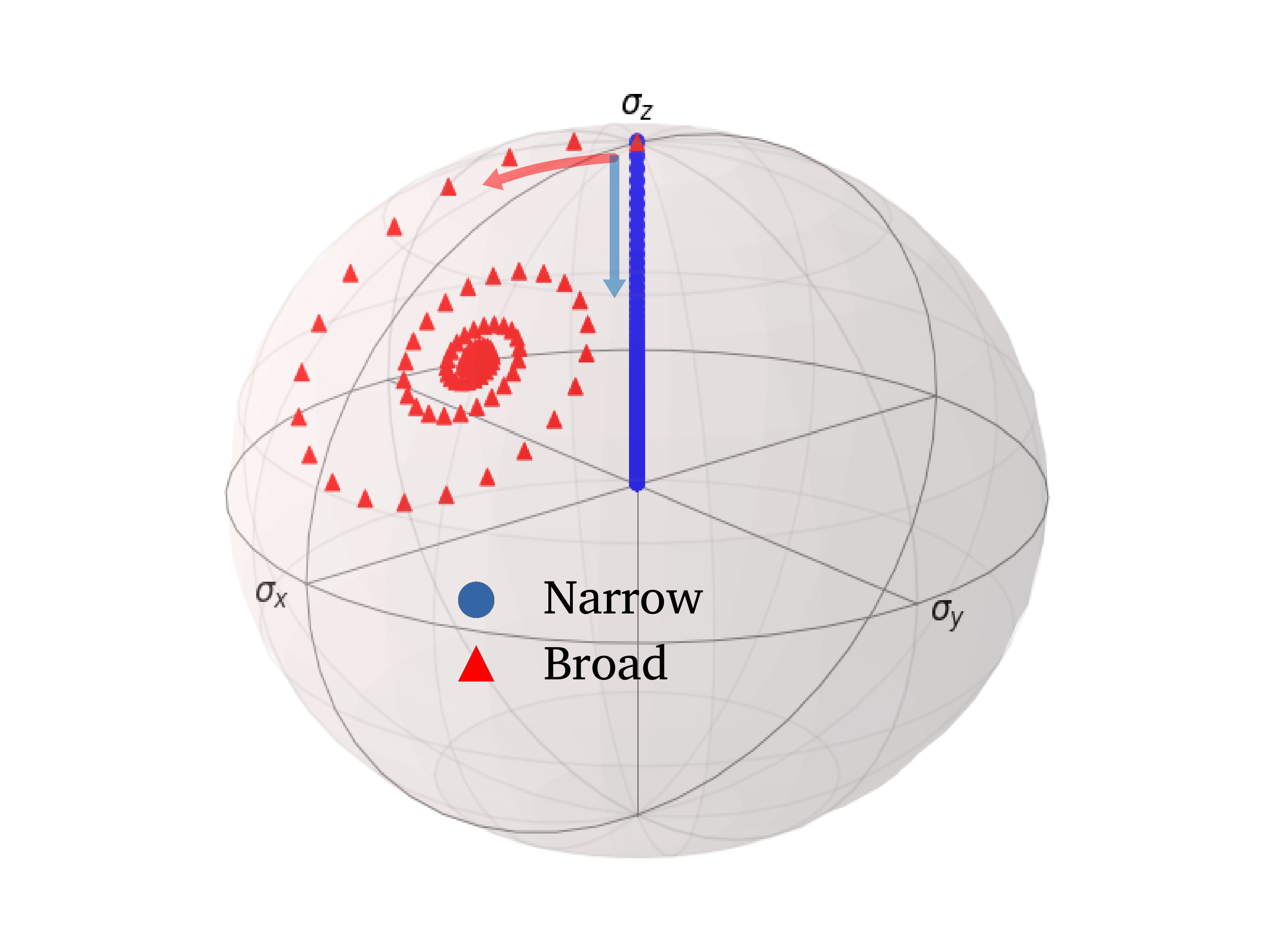}
\includegraphics[width=9cm]{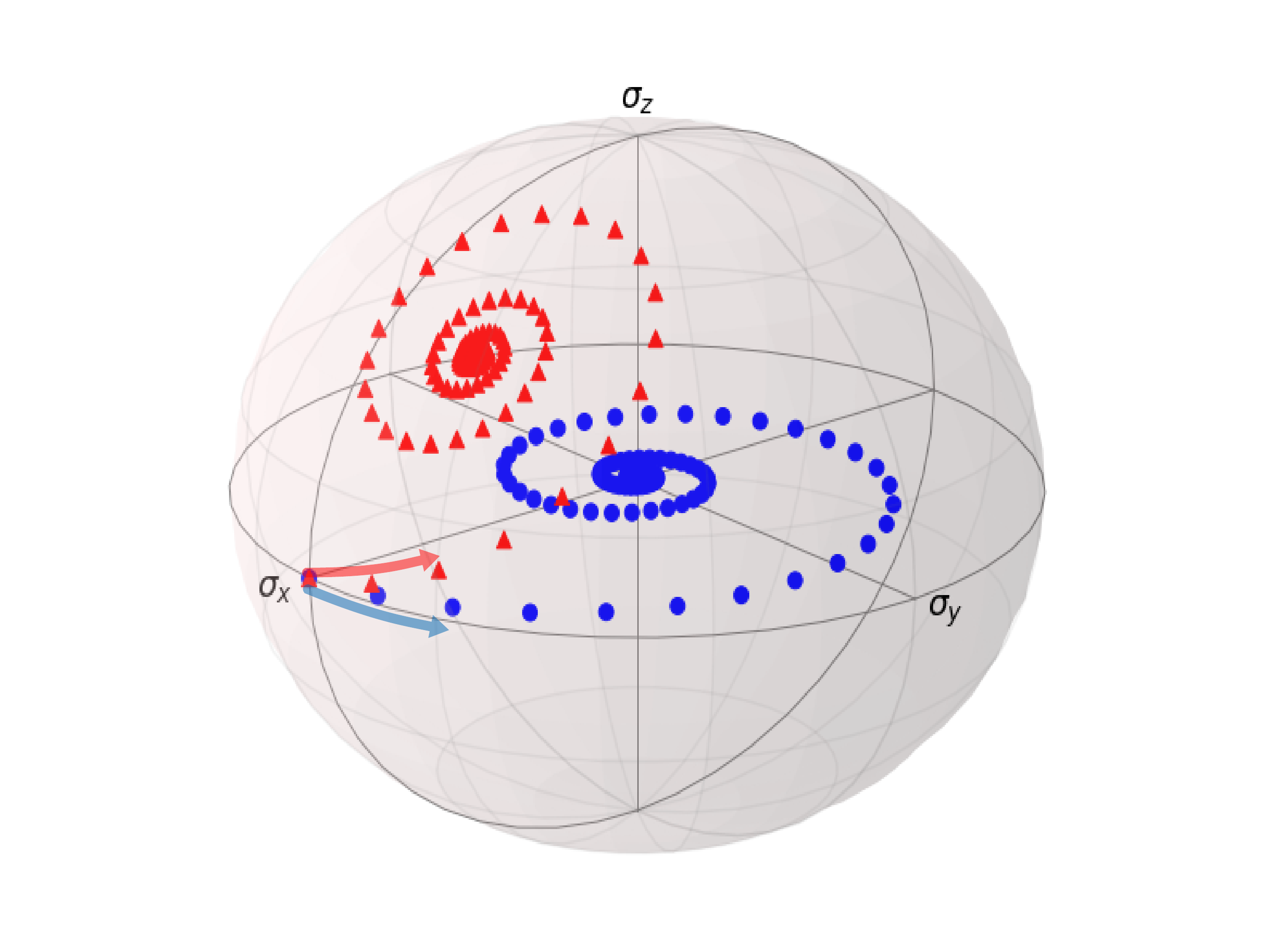}

\caption{ 
Bloch sphere representation of the state $\hatn{\rho}^{(n)}_Y$ under repeated collisions ($n=200$) with narrow and broad Gaussian wave-packets (blue circles and red triangles, respectively). The colored arrows show the direction of the evolution of each state in the sphere starting from the initial states, which are pure (i.e. they lie on the surface of the sphere) and given by $(\rho_Y)^{(0)}_{11}=1$ and $(\rho_Y)^{(0)}_{12}=0$ (upper panel) and $(\rho_Y)^{(0)}_{11}= (\rho_Y)^{(0)}_{12}= 0.5$ (lower panel). All parameters are identical in the two panels, $\Delta_Y=m=1$, $p_0=10$, except the widths, which are $\sigma-\frac{m\Delta_Y}{2p_0} =-0.04$ for narrow wave-packets and $\sigma-\frac{m\Delta_Y}{2p_0} = 0.95$ for broad wave-packets.}
\label{fig-bloch}
\end{figure}

\subsection{Thermalization}

In this section, we explore the dynamics of system $Y$ under repeated collisions with narrow wave-packets as described in section \ref{sec.ensemble}. In particular, we aim to illustrate the thermalization property observed with $\mathbb{S}_{j'j'}^{jj}$ in Eq.~(\ref{S_Y-TR}) when $\mu=\mu_{\rm eff}$. 

\begin{figure}[htb]
\includegraphics[width=7.5cm]{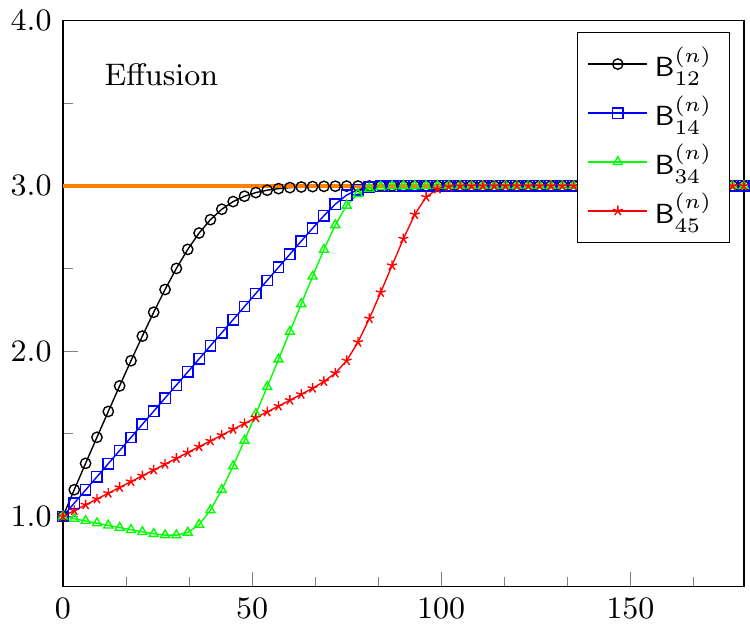}
\includegraphics[width=7.5cm]{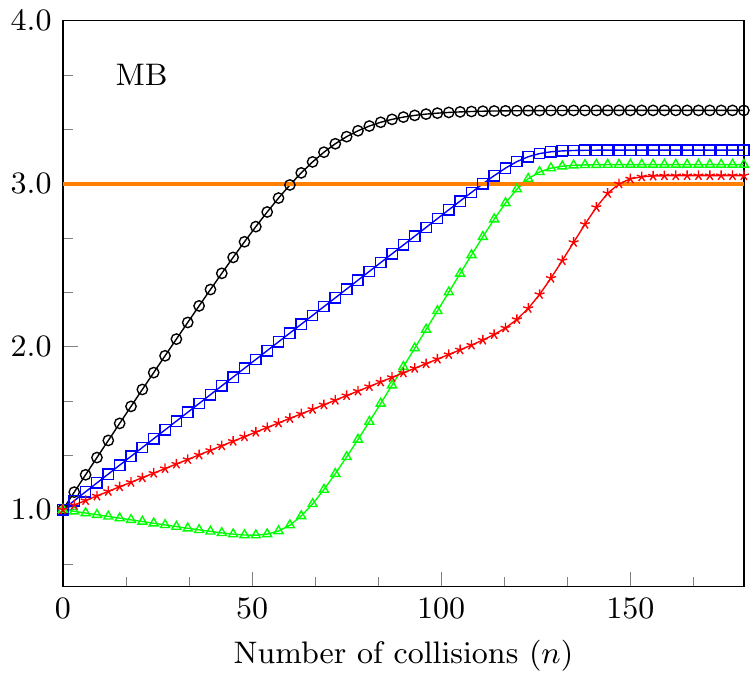}

\caption{\label{fig.therm} Plot of $\mathsf{B}^{(n)}_{jk}$ defined in \eqref{TheBs} computed for an ensemble of narrow Gaussian wave-packets weighted by the effusion distribution in \eqref{effdist.eq} and the Maxwell-Boltzmann (MB) distribution (upper and lower panel, respectively). In both cases, the initial state is the thermal state $e^{-\beta' H_Y}/Z_Y$ with $\beta'=1$ and the mass of the wave-packet is $m=0.5$. The orange line corresponds to $\beta=3$. For clarity, not all pairs of indices $(j,k)$ are plotted.}
\label{fig-bloch}
\end{figure}

For that purpose, we take a system $Y$  of dimension 5 and energy spectrum is $\{e_j=j^2\}_{j=1}^5$, which has non-degenerate Bohr frequencies. As coupling matrix $\mathbb{V}$, we take
\begin{equation}
\mathbb{V}
=\left(\begin{array}{ccccc}
0&1&0&1&0\\
1&0&1&0&1\\
0&1&0&1&0\\
1&0&1&0&1\\
0&1&0&1&0
\end{array}\right) \; .
\end{equation}
To check whether the collisions induce the thermalization of system $Y$, we
evaluate the quantities
\begin{equation}
\label{TheBs}
\mathsf{B}^{(n)}_{jk}\equiv-\frac{1}{e_j-e_k}\ln\frac{(\hatn{\rho}^{(n)}_Y)_{jj}}{(\hatn{\rho}^{(n)}_Y)_{kk}}
\end{equation}
where $\hatn{\rho}^{(n)}_Y=\mathbb{S}^n\hatn{\rho}^{(0)}_Y$ and $\hatn{\rho}^{(0)}_Y$ is the initial state. Since we are interested in the evolution of the populations, the free evolution of system $Y$ between collisions does not play a role. If as $n$ grows, the system $Y$ approaches a thermal distribution $e^{-\beta H_Y}/Z_Y$, then all $\mathsf{B}^{(n)}_{jk}$ converge to $\beta$. We see in Fig.~\ref{fig.therm} that, when the 
wave-packets are weighted with the effusion distribution $\mu_{\rm eff}$ in Eq.~\eqref{effdist.eq}, the system $Y$ thermalizes (upper panel). On the other hand, if they are weighted with the Maxwell-Boltzmann distribution $\mu \propto \exp(-\beta p^2 / 2m)$ (lower panel), the stationary state is not thermal. The evolution of coherences were also tracked (but not depicted) and observed to decay exponentially, as predicted.


\subsection{Ensemble of broad wave-packets}

To complement our analysis, we consider here a mixture of broad wave-packets, each being weighted according to the effusion distribution $\mu_{\rm eff}$. Despite the fact that incident particles have the same distribution of velocities as classical particles effusing from a gas at equilibrium, we will see that the system $Y$ does not thermalize. To illustrate this, we consider again a two-level system $Y$. The condition for a narrow packet for incident momentum $p_0$ is Eq.~(\ref{narrow_cond_gauss}). Since the effusion distribution has its maximum at $p_{\rm max}=\sqrt{m/\beta}$, to satisfy the condition of broad wave-packet for most of the incident particles we set the width of the packet as $\sigma>\Delta_{Y} \sqrt{m\beta}/2$.

\begin{figure}[htb]
\includegraphics[width=8cm]{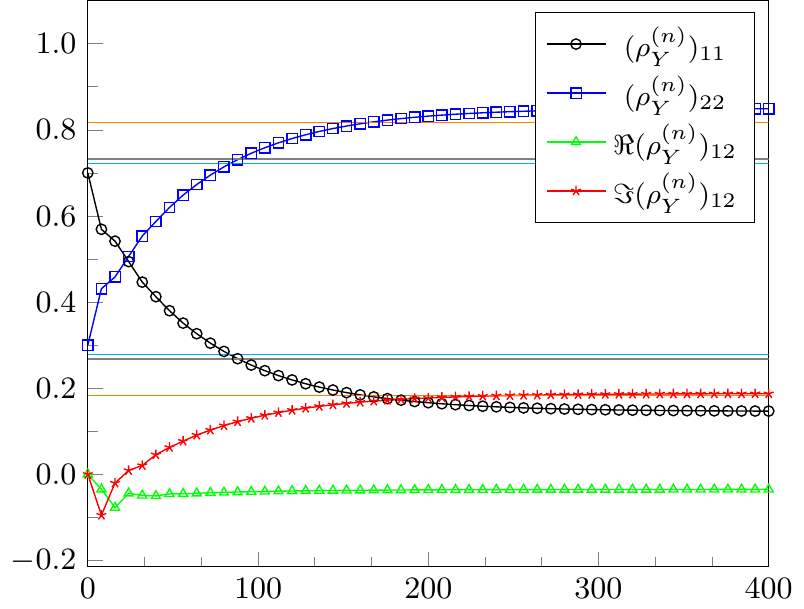}
\includegraphics[width=8cm]{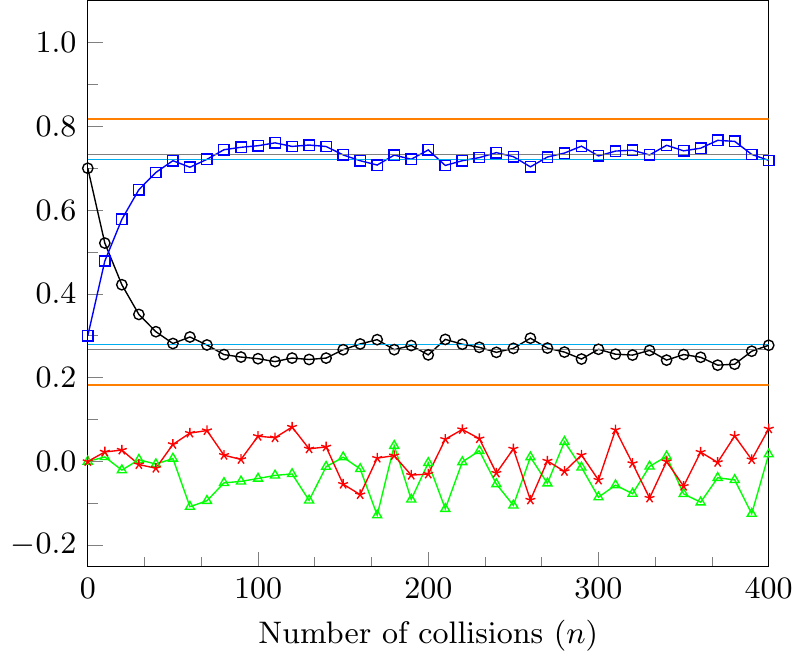}

\caption{\label{thermal-coh-fig} Evolution of $\hatn{\rho}_Y^{(n)}$ when the state of $X$ is a mixture of broad Gaussian wave-packets weighted with the effusion distribution. Upper panel: The values $(\rho_Y)^{(n)}_{ij}$ are obtained by composing the map $\mathbb{S}_Y$. In black we depict $(\rho_Y)^{(n)}_{11}$ and $(\rho_Y)^{(n)}_{22}$ in blue, while the real and imaginary parts of $(\rho_Y)^{(n)}_{12}$ are depicted in green and red, respectively.  Lower panel: The iterated map is given in Eq.~(\ref{concatenation}) with random Poissonian times $\tau_i$. In both panels: The orange lines indicates the thermal values $(\rho_Y)_{11}$ and $(\rho_Y)_{22}$ with inverse temperature $\beta$. The light blue lines indicate the thermal values $(\rho_Y)_{11}$ and $(\rho_Y)_{22}$ with inverse temperature $\beta_C$. The gray lines indicate the values $(\rho_Y)_{11}$ and $(\rho_Y)_{22}$ computed by considering the master equation for the populations ruled by $\mathbb{W}_{ik}=\mathbb{S}_{ii}^{kk}$ as given in Eq.~(\ref{thm-coh.EQ}). The initial state is $(\rho_Y)^{(0)}_{11}=0.3$ and $(\rho_Y)^{(0)}_{12}=0$. The remaining parameters are $m=0.5$, $\sigma=0.31,\beta=3, e_2=2.5,e_1=2$ and $\Delta_Y=0.5$.}
\label{fig-bloch}
\end{figure}

In the upper panel of Fig.~\ref{thermal-coh-fig}, we plot the iteration of $\mathbb{S}$ and observe that, despite the thermal distribution of the average velocities of the wave-packets, the asymptotic state of $Y$ is not thermal, and in fact develops coherences that were absent in the initial state. Notice that these coherences evolve between collisions under the unitary evolution given by the free Hamiltonian $H_{Y}$. Therefore, in contrast to what happens with narrow wave-packets, the time between collisions affects the state of the system. In the lower panel of 
Fig.~\ref{thermal-coh-fig}, we consider the iteration in Eq.~(\ref{concatenation}) with the times $\tau_i$ drawn from a random distribution with Poissonian statistics. This randomness dephases the state of the system, but the populations do not approach their thermal value (orange straight lines). Instead, they fluctuate near a value determined by the stationary state of the map for the populations, i.e., 
the master equation with $\mathbb{W}_{j'j}=\mathbb{S}_{j'j'}^{jj}$ where 
\begin{equation}
\mathbb{S}_{j'j'}^{jj}=\int dp \rho_X(p,p)2P_{j'j}(E_{p} + e_j).
\label{thm-coh.EQ}
\end{equation}
Here $P_{j'j}(E)$ are the transition probabilities defined in Eq.~(\ref{defP}) and $ \rho_X(p,p)$ is the diagonal part of the density matrix of gaussian wave-packets Eq.~\eqref{gaussiandistribution} weighted with the effusion distribution $\mu_{\rm eff}$ Eq.~\eqref{effdist.eq}. A cumbersome but straightforward calculation yields 
%
%
%
\begin{equation}\label{diagbroad}
\rho_X(p,p)=
\frac{\beta_C }{m}\left[\frac{\sigma}{\sqrt{2\pi}} e^{ - \frac{p^2}{2 \sigma^2} }+ \frac{ p}{2\sqrt{r}}\, \text{erf} \Big( \frac{p}{\sqrt{2r} \sigma} \Big) e^{-\beta_C \, \frac{p^2 }{2m} } \right] , 
\end{equation}
with
\begin{equation}
r=1+\frac{\beta\sigma^2}{m},\quad \beta_C=\frac{\beta}{r}.
\label{param-coh.eq}
\end{equation}
The map for the populations given by Eq.~\eqref{thm-coh.EQ} does not satisfy detailed balance if $\sigma$ is not 
negligible, implying that the system does not thermalize when bombarded by these broad packets. It is interesting to notice that the diagonal part \eqref{diagbroad} behaves like an effusion distribution with a temperature $\beta_C$ for $p\gg\sigma$. As shown in the lower plate of Fig.~\ref{thermal-coh-fig}, for Poissonian collision times the system reaches an effective temperature close to $\beta_C$.

Notice, however, that the broad/narrow packet condition depends on $\Delta_Y$, that is, on the energy spectrum of the system. Hence, the same ensemble of packets used for Fig.~\ref{thermal-coh-fig}  will thermalize the system to the temperature of the ensemble $\beta$ if the level spacing $\Delta_Y$ is large enough. One can even induce a crossover from narrow to broad wave-packets by decreasing $\Delta_Y$.

\section{Conclusions}
\label{sec.conclu}

We considered a quantum scattering process between a massive particle $X$ described by a wave-packet and a static system $Y$ and studied the resulting quantum map on $Y$. 
We found that the properties of the map strongly depend on the properties of the wave-packet. 
For wave-packets whose energy width is smaller than the smallest energy level spacing (narrow wave-packet), eigenstate populations and coherences decouple from each other if $Y$ is non-degenerate and the latter decay. Instead, for broad wave-packets, populations and coherences couple and influence each other.
Our central finding is that thermal ensembles of wave-packets, i.e. narrow wave-packets distributed with the effusion probability distribution function, induce decoherence and thermalization in $Y$.

Our results strongly suggest that the distinction between narrow and broad packets could be observable in certain situations, like the interaction between single atoms and a single electromagnetic mode in cavity-QED experiments. Since the broadness of the packet depends on the level spacing $\Delta_Y$, one could tune $\Delta_Y$ to measure the width of the packets and explore the wave-particle duality of atoms and molecules escaping by effusion from a gas in thermal equilibrium. 

The scattering framework that we propose here is very rich and opens many interesting perspectives for the future. 
We are particularly interested in using it as a basis for a quantum thermodynamics formulation. 
Indeed, our present scattering approach avoids many difficulties encountered using other formulations as it is formally exact, autonomous, and the interaction energy is naturally vanishing before and after the collisions.



\acknowledgments{
SLJ is supported by the Doctoral Training Unit on Materials for Sensing and Energy Harvesting (MASSENA) with the grant: FNR PRIDE/15/10935404.
ME is also funded by the European Research Council (project NanoThermo, ERC-2015-CoG Agreement No. 681456). 
F. B. thanks Fondecyt project 1191441 and the Millennium Nucleus ``Physics of active matter'' of the Millennium Scientific Initiative.
Part of this work was conducted at the KITP, a facility supported by the US National Science Foundation under Grant No. NSF PHY-1748958. JMRP acknowledges financial support from the Spanish Government (Grant Contract, FIS-2017-83706-R) and from the Foundational Questions Institute Fund, a donor advised fund of Silicon Valley Community Foundation (Grant number FQXi-IAF19-01).
}

\appendix


\section{Scattering theory} 
\label{sec:appendixscatt}

The main goal of scattering theory is to obtain the scattering operator $\hatn{S}$ which is a one-to-one map between free (incoming) states before the collision to free (outgoing) states after the collision. Whether or not such an operator exists depends on the dynamics during the collision through the interaction potential, which in general can support states which are either free or bound to the interaction region for very long times. However, its existence is guaranteed for a large class of potentials $V(x)$ which vanish fast enough at infinity. In this case, the Hilbert space $\mathcal{H}$ of the full system can be expressed as a direct sum of two mutually orthogonal subspaces $\mathcal{H}=\mathcal{S} \oplus \mathcal{B}$ where $\mathcal{S}$ and $\mathcal{B}$ are the subspaces of scattering and bound states~\cite{Taylor2006,Pearson1988,Yafaev1992}.

According to their definition \eqref{mollerop}, the M{\o}ller operators are isometries of $\mathcal{H}$, i.e., they map state vectors onto the subset of scattering states $\hatn{\Omega}_{\pm}: \mathcal{H} \mapsto \mathcal{S}$ while preserving the norm 
$\braket{\psi|\hatn{\Omega}_{\pm}^{\dagger} \hatn{\Omega}_{\pm}|\psi} = \braket{\psi|\psi}, \forall \ket{\psi} \in \mathcal{H}$. 
The isometric property reads $\hatn{\Omega}_{\pm}^{\dagger} \hatn{\Omega}_{\pm} = \mathbb{I}$. Note that a unitary operator is necessarily an isometry, but the reverse is not true. For instance, here the ``inverse'' M{\o}ller operator $ \hatn{\Omega}^{\dagger}_{\pm}$ acts only on scattering states $\hatn{\Omega}^{\dagger}_{\pm}: \mathcal{S} \mapsto \mathcal{H}$, 
and one has $\hatn{\Omega}_{\pm} \hatn{\Omega}_{\pm}^{\dagger} = \mathbb{I}-\hatn{P}_{\mathcal{B}}$, with $\hatn{P}_{\mathcal{B}}$ the projector onto the space $\mathcal{B}$ of bound states. 
The M{\o}ller operators define the scattering operator according to Eq.~\eqref{scattop} in the main text. Its unitarity follows from the fact that $\hatn{\Omega}_\pm$ have the same range $\mathcal{S}$, a property called asymptotic completeness~\cite{Taylor2006,Pearson1988,Yafaev1992}. 

Some relevant properties of these operators are derived in the next sections.

\subsection{Intertwinning relation}
\label{sec:appendixintertwining}

One of the main properties of  the M{\o}ller operators $\hatn{\Omega}_{\pm}$ is the so-called  intertwining relation:
	\begin{equation}
	\label{intertwining}
	\hatn{H}\hatn{\Omega}_{\pm}=\hatn{\Omega}_{\pm}\hatn{H}_0 \; .
	\end{equation}
The proof follows from the definition of the M{\o}ller operators in Eq. \eqref{mollerop}
	\begin{align}
	e^{ {i \hatn{H}\tau}/{\hbar} } \,\hatn{\Omega}_{\pm} 
	& =  e^{ {i \hatn{H}\tau}/{\hbar} }  \left[ \lim_{t \rightarrow \mp\infty}
	e^{ {i \hatn{H}t}/{\hbar} } e^{ {-i \hatn{H}_0 t}/{\hbar} } \right] \nonumber 	\\
	& =  \lim_{t \rightarrow \mp\infty} 
	e^{ {i \hatn{H}(t+\tau)}/{\hbar} } e^{ {-i \hatn{H}_{0}(t+\tau)}/{\hbar} }e^{ {i \hatn{H}_{0}\tau}/{\hbar} }
	 \nonumber \\
	 &  = \hatn{\Omega}_{\pm} \,e^{ {i \hatn{H}_{0}\tau}/{\hbar} }.
	\end{align}
	Differentiating with respect to $\tau$ at $\tau=0$, we obtain the desired result \eqref{intertwining}. By writing Eq.~\eqref{intertwining} as $\hatn{\Omega}_{\pm}^{\dagger} \hatn{H} \hatn{\Omega}_{\pm} = \hatn{H}_0$, it is clear that such a transformation of the Hamiltonian $\hatn{H}$ is not unitary, since it relates the full Hamiltonian containing the interaction with the free Hamiltonian, which has different energy spectrum. 

From the intertwinning relation, we can straightforwardly derive the important commutation property of the scattering operator
\begin{align}
\hatn{S}\hatn{H}_0&=\hatn{\Omega}_{-}^{\dagger}\hatn{\Omega}_{+}\hatn{H}_0=\hatn{\Omega}_{-}^{\dagger}\hatn{H}\hatn{\Omega}_{+}
\nonumber \\ 
&=\hatn{H}_0\hatn{\Omega}_{-}^{\dagger}\hatn{\Omega}_{+}=\hatn{H}_0\hatn{S}\quad \Rightarrow\quad [\hatn{S},\hatn{H}_0]=0,
\end{align}
which implies the conservation of the total energy  in the scattering event and that  
 the elements of the scattering operator in the eigenbasis of $H_{0}$, i.e. $\bra{p',j'}\hatn{S}\ket{p,j}$, are proportional to $\delta(E_{p}+e_{j}-E_{p'}-e_{j'})=\delta(E_{p}-E_{p'}-\Delta_{j'j})$, allowing us to write these terms as in Eq.~\eqref{expresionS}.
	
Another important consequence of the interwinning relation is that, if $\ket{p,j}$ is an improper eigenstate of $H_{0}$ with energy $E=E_{p}+e_{j}=p^{2}/(2m)+e_{j}$, then $\Omega_\pm \ket{p,j}$ is an  eigenstate of the full Hamiltonian $H$ with the same energy:
\begin{equation}\label{scatphi0}
H\Omega_{\pm} \ket{p,j}=\Omega_{\pm} H_{0}\ket{p,j}=E\,\Omega_{\pm}\ket{p,j}.
\end{equation}

\subsection{The $T$ operator}
\label{sec:appendixtoperator}

A crucial tool in scattering theory are the resolvents or Green's operators associated to the free and the full Hamiltonian:
\begin{equation}
G_{0}(z)\equiv(z-H_{0})^{-1} \, ; \qquad G(z)\equiv(z-H)^{-1}
\end{equation}
defined for any complex number $z$ which does not belong to the spectrum of $H_{0}$ and $H$, respectively. It is not hard to prove that these operators verify
\begin{equation}\label{gg0}
\begin{split}
G(z)&=G_{0}(z)+G_{0}(z)VG(z) \\
&=G_{0}(z)+G(z)VG_{0}(z).
\end{split}
\end{equation}
We also define the $T$ operator as
\begin{equation}
T(z)\equiv V+VG(z)V.
\end{equation}
Applying $G_{0}(z)$ to this equation and making use of  Eq.~\eqref{gg0}, we get
\begin{equation}\label{g0tgv}
G_{0}(z)T(z)=G(z)V
\end{equation}
which relates the two resolvents through the $T$ operator and the interaction potential.

To relate the scattering and the $T$ operator, we rewrite the M{\o}ller operators \eqref{mollerop} as
 	\begin{align}
\hatn{\Omega}_{\pm} 
=& 1 + \int^{\mp \infty}_{0} dt \, \frac{d}{dt} (e^{i \hatn{H} t/{\hbar}}\,e^{-i \hatn{H}_0 t/{\hbar}}) \nonumber\\
=& - \lim_{\epsilon \rightarrow 0^+} \big \lbrack e^{\pm\epsilon t/\hbar} e^{i \hatn{H} t/{\hbar}}\,e^{-i \hatn{H}_0 t/{\hbar}} \big \rbrack^{\mp \infty}_{0} \nonumber \\
& + \lim_{\epsilon \rightarrow 0^+} \int^{\mp \infty}_{0} dt \,  e^{\pm\epsilon t/\hbar} \frac{d}{dt} (e^{i \hatn{H} t/{\hbar}}\,e^{-i \hatn{H}_0 t/{\hbar}}) \nonumber \\
=& \lim_{\epsilon \rightarrow 0^+} \mp\frac{\epsilon}{\hbar} \int^{\mp \infty}_{0} dt\, e^{\pm\epsilon t/\hbar}e^{i \hatn{H} t/{\hbar}}\,e^{-i \hatn{H}_0 t/{\hbar}} \; , \label{mollerint}
	\end{align}
where integration by parts was used from the second to the third equality.
Its action on the improper eigenstates of $\hatn{H}_0$ gives
	\begin{align}
	\hatn{\Omega}_{\pm} \ket{p,j}  &=\lim_{\epsilon \rightarrow 0^+} \mp\frac{\epsilon}{\hbar} \int^{\mp \infty}_{0} dt\, e^{-i(E\pm i\epsilon) t/\hbar}\,e^{i \hatn{H} t/{\hbar}}\,\ket{p,j}
	 \nonumber \\ &=\lim_{\epsilon \rightarrow 0^+} \pm\, i\, \epsilon\, \hatn{G}(E \pm i \epsilon) \ket{p,j} \nonumber \\ &= \ket{p,j} +  \hatn{G}(E\pm i 0) \hatn{V} \ket{p,j}
	,\label{mollerresolvent}
	\end{align}
where $E=E_{p}+e_{j}=p^2/(2m)+e_{j}$ is the energy  of  the state $\ket{p,j}$ and we have introduced the notation $f(z+i0)\equiv\lim_{\epsilon \rightarrow 0^+}f(z+i\epsilon)$. The second equality follows from a direct calculation of the integral in Eq.~\eqref{mollerint}, while the third is obtained using Eq.~\eqref{gg0} and $G_{0}(z)\ket{p,j}=(z-E)^{-1}\ket{p,j}$.
Multiplying Eq.~(\ref{mollerresolvent}) by $V$ we get the useful expression
\begin{equation}\label{vot}
V\hatn{\Omega}_{\pm} \ket{p,j}=T(E\pm i0)\ket{p,j}.
\end{equation}
Recalling the representation of the Dirac delta, $\pi\delta(x)=\lim_{\epsilon\to 0^{+}}\,{\rm Im} (x-i\epsilon)^{-1}$, one can prove that Eq.~(\ref{mollerresolvent})  implies the identity 
\begin{align}
(\hatn{\Omega}_+ -\hatn{\Omega}_-)\ket{p,j} & = [\hatn{G}(E+i0)-\hatn{G}(E-i0)]\hatn{V}\ket{p,j} \nonumber \\
& = -2\pi i\delta(E-\hatn{H})\hatn{V}\ket{p,j} \; .
\label{magic:ec-a}
\end{align}
The action of the scattering operator on an eigenstate of $\hatn{H}_0$ can now be computed from its definition \eqref{scattop} and Eq.~\eqref{magic:ec-a}:
\begin{align}
\hatn{S}\ket{p,j} \nonumber & = \hatn{\Omega}_{-}^{\dagger}\hatn{\Omega}_{+}\ket{p,j} \\ 
& = (1+ \hatn{\Omega}_{-}^{\dagger} (\hatn{\Omega}_{+} - \hatn{\Omega}_{-}))\ket{p,j} \nonumber \\
& = \ket{p,j} -2\pi i \,\hatn{\Omega}_{-}^{\dagger} \delta(E-\hatn{H}) \hatn{V} \ket{p,j} \; .
\end{align}
Multiplying by $\bra{p',j'}$ and taking into account that $\Omega_{-}\ket{p',j'}$ is an eigenstate of $H$ with eigenvalue $E_{{p'}}+e_{j'}$:
\begin{align}
\bra{p',j'}\hatn{S}\ket{p,j} \nonumber & = \delta_{j'j} \delta(p'-p) \\ & - 2\pi i \delta(E_{p}-E_{p'}-\Delta_{j'j})\bra{p',j'}\hatn{\Omega}_{-}^{\dagger}  \hatn{V} \ket{p,j} \; .
\label{scatopp14}
\end{align}
We can eliminate the M{\o}ller operator using Eq.~\eqref{vot} and the property $T^{\dagger}(z)=T(z^*)$, yielding $\bra{p',j'}\hatn{\Omega}_{-}^{\dagger}  \hatn{V} \ket{p,j} = \bra{p',j'} T(E+ i0)\ket{p,j}$. Inserting this last expression into \eqref{scatopp14}, one finally gets
%

\begin{widetext}
\begin{align}
\label{scattopdecomposition2}
\bra{p',j'}\hatn{S}\ket{p,j} & = \delta_{j'j}\delta(p'-p) - 2 \pi i \delta(E_{p}-E_{p'}-\Delta _{j'j}) \times \braket{p',j'|\hatn{T}(E+ i0)|p,j} \nonumber \\
& = \delta(E_{p}-E_{p'}-\Delta_{j'j})\left[\delta_{j'j}\delta_{\alpha'\alpha}\frac{|p|}{m}-2\pi i\braket{p',j'|\hatn{T}(E+ i0)|p,j} \right] \; .
\end{align}
Comparing this expression with Eq.~\eqref{expresionS} and taking into account that $j'=j$ implies $|p'|=|p|$, we get
\begin{equation}\label{smatrixt}
s_{j'j}^{(\alpha' \alpha)}(E)=\delta_{j'j}\delta_{\alpha'\alpha}-\frac{2\pi i \,m}{\sqrt{|pp'|}}\braket{p',j'|\hatn{T}(E+ i0)|p,j} .
\end{equation}
In this way, we have related the entries of the scattering matrix  $\bf s$ to the $T$ operator. In the next section of this appendix, we relate the elements of the $T$ operator to the transmission and reflection coefficients which, in turn, can be computed by solving the stationary Schr\"odinger equation. 

\subsection{Scattering states}
\label{sec:appendixscattstates}

The previous definitions and relationships can be used to compute the $T$ operator and the scattering operator $S$ in specific situations. For this purpose, let us introduce the so-called scattering states $\ket{p,j}_{+}$ defined as
\begin{align}
\label{scatdefpsi}
\ket{p,j}_{+}\equiv \Omega_{+}\ket{p,j} &=\ket{p,j} +  \hatn{G}(E+ i 0) \hatn{V} \ket{p,j}\nonumber \\
&=\ket{p,j} +  \hatn{G}_{0}(E+ i 0)T(E+ i 0) \ket{p,j} \; .
\end{align}
Here, we have used Eqs.~\eqref{mollerresolvent} and \eqref{g0tgv}.
We also use the following identity, derived from \eqref{vot} and \eqref{scatdefpsi}:
\begin{equation}\label{scatg0v}
\ket{p,j}_{+}=\ket{p,j} +  \hatn{G}_{0}(E+ i 0) \hatn{V} \ket{p,j}_{+}.
\end{equation}
The scattering state $\ket{p,j}_{+} 
$ is an eigenstate of $H$ with energy $E=E_{p}+e_{j}$ as shown in Eq.~\eqref{scatphi0}. Moreover, the asymptotic behavior of its wave function in real space $\braket{x|p,j}_{+} 
$ contains all the necessary information to calculate the operators $T$ and $S$.
To prove this, we start by expressing Eq.~\eqref{scatg0v} in the position representation, using the plane-wave introduced in \eqref{planewavex}:
\begin{align}
\braket{x|p,j}_{+} &=\frac{e^{ikx}}{ \sqrt{2 \pi \hbar}}\,\ket{j}+\int_{-\infty}^{{\infty}} dx' \int_{{-\infty}}^{{\infty}} dx''\braket{x|G_0(E+i0)|x'}\braket{x'|V| x''}\braket{x''|p,j}_{+} \nonumber \\
&=\frac{e^{ikx}}{ \sqrt{2 \pi \hbar}}\,\ket{j}+\int_{-\infty}^{{\infty}} dx'\, V(x')\braket{x|G_0(E+i0)|x'}
\nu\braket{x'|p,j}_{+} 
\label{LS-p:ec}
\end{align}
with $k=p/\hbar$. The resolvent  in position representation $G^{+}_{0}(x,x')\equiv \braket{x|G_0(E+i0)|x'}$ is an operator in ${\cal H}_{Y}$ that obeys the equation
\begin{equation}
\left[E- \hatn{H}_Y+\frac{\hbar^2}{2m}\frac{d^2}{dx^2}\right]G^{+}_0(x,x')=
\delta(x-x')\otimes {\mathbb I}_{Y} \; .
\end{equation}
The solution is
\begin{equation}
\label{green0:ec}
G^{+}_0(x,x')=\sum_{j'}\frac{m}{ik_{j'}\hbar^2}e^{ik_{j'}|x-x'|}\ket{{j'}}\bra{{j'}} \; ,
\end{equation} 
with $k_{j'}=\sqrt{{2m}(E-e_{j'})}/\hbar$,
as one can check by direct substitution.
Inserting this expression  in Eq.~\eqref{LS-p:ec}, we get
\begin{equation}
\braket{x|p,j}_{+}=\frac{e^{ikx}}{ \sqrt{2 \pi \hbar}}\,\ket{j}+\sum_{{j'}}\frac{m}{ik_{j'}\hbar^2}\ket{{j'}}\int_{-\infty}^{{\infty}} dx' e^{ik_{j'}|x-x'|} V(x')
\bra{j'} \hatn{\nu} \braket{x'|p,j}_{+} .
\end{equation}
We are interested in the asymptotic behavior of $\braket{x|p,j}_{+}\,$ far from the collision region. Consider first the case $p>0$ and $x\to -\infty$. Due to the presence of $V(x')$, only the values of $x'$ in the collision region contribute to the integral. Therefore, we can replace $|x-x'|$ by $x'-x$, yielding
\begin{align}
\label{asymptminusinf}
\braket{x|p,j}_{+}\,\underset{x\to-\infty}{=}
\frac{1}{\sqrt{2 \pi \hbar}}\,\Big[e^{ikx}\,\ket{j}+\sum_{j'} r^L_{{j'}j}\,e^{-ik_{j'}x}\ket{{j'}}\Big] \; ,
\end{align}
where 
\begin{align}
\label{asymptminusinfBis}
r^L_{{j'}j} = \frac{m\,e^{-ik_{j'}x}}{ik_{j'}\hbar^2} \int_{-\infty}^{{\infty}} dx' e^{ik_{j'}x'} V(x') \bra{j'} \hatn{\nu} \braket{x'|p,j}_{+} \;. 
\end{align}

Hence, the scattering state $\ket{p,j}_{+}$ is a superposition of wave-packets with certain amplitudes $r^L_{j'j}$, which we need to relate to $T$ and the $S$ operators. To achieve this, we express Eq.~\eqref{scatdefpsi} in position representation
\begin{align}
\braket{x|p,j}_{+} &=\frac{1}{ \sqrt{2 \pi \hbar}}\left[e^{ikx}\ket{j}+\int_{-\infty}^{{\infty}} dx' \int_{-\infty}^{{\infty}} dx''e^{ikx''} G^+_0(x,x')\braket{x'|T(E+i0)| x'',j}\right] \nonumber \\
&= \frac{1}{ \sqrt{2 \pi \hbar}}\left[e^{ikx}\ket{j}+\sum_{j'}\frac{m}{ik_{j'}\hbar^2}\ket{j'}\int_{-\infty}^{{\infty}} dx' \int_{-\infty}^{{\infty}} dx''e^{ikx''} e^{ik_{j'}|x-x'|}\braket{x',j'|T(E+i0)| x'',j}\right] \; .
\label{LS-p:ec2}
\end{align}
To conform with the asymptotic behavior \eqref{asymptminusinf}, the integrand must be localized in a finite region. We can then replace again $|x-x'|$ by $x'-x$ when $x\to -\infty$, obtaining
\begin{equation}\label{asympt2}
\braket{x|p,j}_{+}\,\underset{x\to-\infty}{=}\frac{1}{ \sqrt{2 \pi \hbar}}\left[e^{ikx}\ket{j}+\sum_{j'}\frac{m e^{-ik_{j'}x}}{ik_{j'}\hbar^2}\ket{j'}\int_{-\infty}^{{\infty}} dx' \int_{-\infty}^{{\infty}} dx''e^{ikx''} e^{ik_{j'}x'}\braket{x',j'|T(E+i0)| x'',j}\right] \; .
\end{equation}
Comparing \eqref{asymptminusinf} and \eqref{asympt2}, we get
\begin{equation}
\label{rt1}
r^L_{{j'}j}=\frac{m}{ik_{j'}\hbar^2}\int_{-\infty}^{{\infty}} dx' \int_{-\infty}^{{\infty}} dx''e^{ikx''} e^{ik_{j'}x'}\braket{x',j'|T(E+i0)| x'',j} \; .
\end{equation}
\end{widetext}
If we go back to the momentum representation taking into account that $\ket{p}= (\sqrt{2\pi\hbar})^{-1}\int_{-\infty}^{{\infty}}dx e^{ikx} \ket{x}$ then equation \eqref{rt1} reduces to:
\begin{equation}\label{rinprep}
r^L_{{j'}j}=-{2\pi i}\,\frac{ m}{p_{j'}}\braket{-p_{j'},j'|T(E+i0)| p,j}
\end{equation}
with $p_{j'}=\hbar k_{j'}=\sqrt{p^2-2m\Delta_{{j'j}}}$.
Introducing the value of $\braket{p',j'|\hatn{T}(E+ i0)|p,j}$ given by Eq.~\eqref{rinprep} in Eq.~\eqref{smatrixt}, we can express the entries of the scattering matrix in terms of the amplitudes $r^L_{j'j}$:
\begin{equation}\label{smatrixt2}
\hat r^L_{{j'}j}\equiv s_{j'j}^{(- +)}(E)=\sqrt{\frac{|p'|}{|p|}}r^L_{{j'}j}
\end{equation}
since the sign of the momentum of the incident wave $\ket{p,j}$ is $\alpha=+$ and the one of the reflected wave $\ket{-p_{j'},j'}$ is $\alpha=-$.

The asymptotic behavior of $\ket{p,j}_{+}$ for $x\to +\infty$ is analyzed in a similar way. First we proceed as above to get the analogous to Eq.~\eqref{asymptminusinf}, which reads (notice that we can now include the first term in Eq.~\eqref{asymptminusinf} in the sum):
\begin{equation}
\braket{x|p,j}_{+}\,\underset{x\to +\infty}{=}
\frac{1}{ \sqrt{2 \pi \hbar}}\,\sum_{j'} t^L_{{j'}j}\,e^{ik_{j'}x}\ket{{j'}} \; .\label{asymptplusinf}
\end{equation}
The position representation of Eq.~\eqref{scatdefpsi} is still given by Eq.~\eqref{LS-p:ec2}, but now the absolute value in the exponential is $|x-x'|=x-x'$, yielding
\begin{widetext}
\begin{equation}\label{asympt2b}
\braket{x|p,j}_{+}\,\underset{x\to\infty}{=}\frac{1}{ \sqrt{2 \pi \hbar}}\left[e^{ikx}\ket{j}+\sum_{j'}\frac{m e^{ik_{j'}x}}{ik_{j'}\hbar^2}\ket{j'}\int_{-\infty}^{{\infty}} dx' \int_{-\infty}^{{\infty}} dx''e^{ikx''} e^{-ik_{j'}x'}\braket{x',j'|T(E+i0)| x'',j}\right]
\end{equation}
\end{widetext}
and therefore
\begin{equation}
t^L_{{j'}j}=\delta_{j'j}-{2\pi i}\,\frac{ m}{p_{j'}}\braket{p_{j'},j'|T(E+i0)| p,j} \; .
\end{equation}
Finally, the corresponding entry in the scattering matrix reads
\begin{equation}\label{smatrixt2c}
\hat t^L_{{j'}j}\equiv s_{j'j}^{(+ +)}(E)=\sqrt{\frac{|p'|}{|p|}}t^L_{{j'}j} \; .
\end{equation}

Summarizing, to obtain the entries of the scattering matrix with $\alpha=+$, one has to solve the stationary Schr\"odinger equation $H\ket{\psi}=E\ket{\psi}$ for an improper state $\ket{\psi}$ whose position representation $\braket{x|\psi}\in {\cal H}_{Y}$ has the following asymptotic behavior:
\begin{equation}
\braket{x|\psi}=\left\{
\begin{array}{ll}
\displaystyle e^{ik_jx}\ket{j}+\sum_{j'} r^L_{j'j}e^{-ik_{j'}x}\ket{j'}  &\quad  \mbox{for $x\to -\infty$}\\ \\
\displaystyle
\sum_{j'} t^L_{j'j}e^{ik_{j'}x}\ket{j'}&\quad \mbox{for $x\to +\infty$}
\end{array}\right.
\label{eq:left-sc}
\end{equation}
The solution is unique and determines the amplitudes $t^L_{j'j}$ and $r^L_{j'j}$ which in turn determine the entries of the scattering matrix $\hat t^L_{j'j}$ and $\hat r^L_{j'j}$, as prescribed in Eqs.~\eqref{smatrixt2} and \eqref{smatrixt2c}.

Repeating the whole analysis with $p<0$, we get identical results. The solution of the Schr\"odinger equation now must obey the asymptotic conditions:
\begin{equation}
\braket{x|\psi}=\left\{
\begin{array}{ll}
\displaystyle \sum_{j'} t^R_{j'j}e^{-ik_{j'}x}\ket{j'}   &\quad  \mbox{for $x\to -\infty$}\\ \\
\displaystyle e^{-ik_jx}\ket{j}+\sum_{j'} r^R_{j'j}e^{ik_{j'}x}\ket{j'}
&\quad \mbox{for $x\to +\infty$}
\end{array}\right.
\label{eq:right-sc}
\end{equation}
and we finally obtain
\begin{align}
\hat t^R_{{j'}j}\equiv s_{j'j}^{(- -)}(E) & = \sqrt{\frac{|p'|}{|p|}}t^R_{{j'}j} \nonumber \\
\hat r^R_{{j'}j}\equiv s_{j'j}^{(+ -)}(E) & = \sqrt{\frac{|p'|}{|p|}}r^R_{{j'}j}\; ,
\end{align}
for transmission and reflection from the right, respectively.

\begin{widetext}

\subsection{The scattering matrix}
\label{sec:appendixsmatrix}

We order the channels such that the first $N_{\rm open}$ are right propagating and the last $N_{\rm open}$ are left propagating. In this way we write
\begin{equation}
\mathbf{s}(E)=\left(
\begin{array}{cc}
\hat{\mathbf{r}}^L & \hat{\mathbf{t}}^R \\
\hat{\mathbf{t}}^L & \hat{\mathbf{r}}^R
\end{array}\right)=
\left(
\begin{array}{cc}
\hatn{\mathbf{s}}^{-+} & \hatn{\mathbf{s}}^{--} \\
\hatn{\mathbf{s}}^{++} & \hatn{\mathbf{s}}^{+-}
\end{array}\right)
\end{equation}
which satisfies the identities
\begin{align}
\mathbf{s}(E)\mathbf{s}^\dag(E) =
\left(
\begin{array}{cc}
\hat{\mathbf{r}}^L \hat{\mathbf{r}}^{L\dag}+\hat{\mathbf{t}}^R\hat{\mathbf{t}}^{R\dag}& \hat{\mathbf{r}}^L \hat{\mathbf{t}}^{L\dag}+ \hat{\mathbf{t}}^R \hat{\mathbf{r}}^{R\dag}\\
\hat{\mathbf{t}}^L \hat{\mathbf{r}}^{L\dag}+\hat{\mathbf{r}}^R\hat{\mathbf{t}}^{R\dag}&\hat{\mathbf{t}}^L \hat{\mathbf{t}}^{L\dag}+\hat{\mathbf{r}}^R \hat{\mathbf{r}}^{R\dag}
\end{array}\right) = \left(
\begin{array}{cc}
\mathbf{1}& \mathbf{0}\\
 \mathbf{0}& \mathbf{1}
\end{array}\right)
\label{prop1}
\end{align}

\begin{align}
\mathbf{s}^\dag(E)\mathbf{s}(E) =
\left(
\begin{array}{cc}
\hat{\mathbf{r}}^{L\dag}\hat{\mathbf{r}}^L + \hat{\mathbf{t}}^{L\dag}\hat{\mathbf{t}}^L&\hat{\mathbf{r}}^{L\dag}\hat{\mathbf{t}}^R+ \hat{\mathbf{t}}^{L\dag}\hat{\mathbf{r}}^R \\
\hat{\mathbf{t}}^{R\dag}\hat{\mathbf{r}}^L +\hat{\mathbf{r}}^{R\dag}\hat{\mathbf{t}}^L&\hat{\mathbf{t}}^{R\dag}\hat{\mathbf{t}}^R +\hat{\mathbf{r}}^{R\dag}\hat{\mathbf{r}}^R 
\end{array}\right) = \left(
\begin{array}{cc}
\mathbf{1}& \mathbf{0}\\
 \mathbf{0}& \mathbf{1}
\end{array}\right).
\label{prop2}
\end{align}
following from the unitarity of the scattering operator $\hatn{S}^\dag \hatn{S}=\mathbb{I}$ and $\hatn{S}\hatn{S}^\dag=\mathbb{I}$, respectively.  
We sketch the proof of \eqref{prop1}. Starting from
\begin{align}
\bra{p',j'}\hatn{S}^\dag \hatn{S}\ket{p,j} = \delta(p'-p)\delta_{j'j} \; ,
\end{align}
inserting the resolution of identity and expressing the Dirac delta as a function of the kinetic energies, we get
\begin{align}
\sum_k\int dp''\bra{p',j'}\hatn{S}^\dag\ket{p'',k}\bra{p'',k} \hatn{S}\ket{p,j} =\frac{|p|}{m}\,\delta(E_{p'}-E_{p})\delta_{j'j}\delta_{\alpha' \alpha}\, .
\end{align}
We now write the matrix elements of the operator $S$ in terms of the scattering matrix $s_{j'j}^{(\alpha'\alpha)}$ using Eq.~\eqref{expresionS} and perform the  integral over $p''$  by recalling that  $dE_{p''}=|p''|dp''/m$:
\begin{equation}
{\delta(E_{p'}-E_{p}-\Delta_{j'j})} \frac{\sqrt{|p\,p'|}}{m}\sum_{k}\sum_{\alpha''=\pm}[s_{kj'}^{(\alpha''\alpha')}(E_{p'}+e_{j'})]^*s_{kj}^{(\alpha''\alpha)}(E_{p}+e_{j})=\frac{|p|}{m}\,\delta(E_{p}-E_{p'})\delta_{j'j}\delta_{\alpha' \alpha} \; .
\label{jefe}
\end{equation}
Taking into account that the delta function in the right hand side of the equation implies $j=j'\Rightarrow |p|=|p'|$, one obtains
\begin{equation}
\sum_{k}\sum_{\alpha''=\pm}[s_{kj}^{(\alpha''\alpha')}(E)]^*s_{kj}^{(\alpha''\alpha)}(E)=\delta_{j'j}\delta_{\alpha' \alpha}
\end{equation}
which is Eq.~(\ref{prop1}) as a careful inspection shows. To prove Eq.~(\ref{prop2}), one can follow similar steps starting from $\bra{p',j'}\hatn{S}\hatn{S}^\dag\ket{p,j}=\delta(p'-p)\delta_{j'j}$.
\end{widetext}

\section{Scattering matrix for a delta potential}
\label{app:ejemplo-delta}

Here we provide explicit computations of $\hat{\mathbf{t}}=\hat{\mathbf{t}}^{L}=\hat{\mathbf{t}}^{R}$ and $\hat{\mathbf{r}}=\hat{\mathbf{r}}^{L}=\hat{\mathbf{r}}^{R}$ for $V(x)=g\delta(x)$.
According to section \ref{sec:appendixscattstates}, we have to solve the stationary Schr\"odinger equation
$H\ket{\psi}=E\ket{\psi}$ which, in position representation, can be written as
\begin{equation}
\label{eq:schro}
\left\{E-\hatn{H}_Y+\frac{\hbar^2}{2m}\frac{d^2}{dx^2}\right\}\sum_k\psi_k(x)\ket{k}=\hatn{V}\sum_k\psi_k(x)\ket{k} \; .
\end{equation}
where $\braket{x|\psi}=\sum_k\psi_k(x)\ket{k}$. We look for solutions obeying the asymptotic boundary conditions Eq.~\eqref{eq:left-sc} or, equivalently, \eqref{eq:right-sc}.
Since the support of the potential is a single point at $x=0$, the asymptotic conditions are fulfilled for every $x\neq 0$:
\begin{equation}
\braket{x|\psi}=\left\{
\begin{array}{ll}
\displaystyle e^{ik_jx}\ket{j}+\sum_{j'} r_{j'j}e^{-ik_{j'}x}\ket{j'}  &\quad  \mbox{for $x<0$}\\ \\
\displaystyle
\sum_{j'} t_{j'j}e^{ik_{j'}x}\ket{j'}&\quad \mbox{for $x>0$.}
\end{array}\right.
\label{eq:left-sc-delta}
\end{equation}
Projecting the Schr\"odinger equation \eqref{eq:schro} onto $\bra{j'}$, we get
\begin{equation}\label{eq:schro2}
(E - e_{j'})\psi_{j'}(x)+\frac{\hbar^2}{2m}\psi_{j'}''(x) = g\delta(x)\sum_k\psi_k(x) \bra{j'}\hatn{\nu}\ket{k} \; .
\end{equation}
The wave functions $\psi_{j'}(x)$ must be everywhere continuous, while their first derivative $\psi'_{j'}(x)$ has a jump discontinuity at $x=0$. Imposing the continuity condition to the solution Eq.~(\ref{eq:left-sc-delta}), we get $\delta_{j'j}+r_{j'j}=t_{j'j}$ or, in matrix form
\begin{equation}
\mathbb{I}+\mathbf{r}= \mathbf{t}
\label{eq:cont}
\end{equation}

To obtain the value of the jump discontinuity,
we integrate both sides of Eq.~(\ref{eq:schro2}) in an interval $[-\epsilon,\epsilon]$ and then take $\epsilon\to 0^{+}$:
\begin{equation}
\frac{\hbar^2}{2m}[\psi_{j'}'(0^+)-\psi_{j'}'(0^-)] =g\sum_k\psi_k(0) \bra{{j'}}\hatn{\nu}\ket{k} \; .
\label{eq:jump}
\end{equation}

According to \eqref{eq:left-sc-delta}, the derivatives at $x=0$ are $\psi_{j'}'(0^+)=ik_{j'}t_{j'j}$ and $\psi_{j'}'(0^-)=ik_{j'}(\delta_{j'j}-r_{j'j})=ik_{j'}(2\delta_{j'j}-t_{j'j})$, where in the last equality we have used Eq.~(\ref{eq:cont}). From continuity we also have $\psi_{k}(0)=t_{kj}$. Collecting these results, Eq.~(\ref{eq:jump}) yields
\[
\frac{i \hbar^2}{m}k_{j'}[t_{j'j}-\delta_{j'j}]=g\sum_k t_{kj} \bra{j'}\hatn{\nu}\ket{k} \; .
\]
This relation can be expressed in matrix form by defining $\mathbb{D}$ as the diagonal matrix with elements $\mathbb{D}_{jj}=k_j$ and $\mathbb{V}$ the matrix with elements $\mathbb{V}_{jk}=\bra{j}\hatn{\nu}\ket{k}$
\[
\mathbb{D}[\mathbf{t}-\mathbb{I}]=\frac{mg}{\hbar^2 i}\mathbb{V}\,\mathbf{t}
\]
hence 
\begin{equation}
\mathbf{t}=[\,\mathbb{I}+\frac{img}{\hbar^2}\,\mathbb{D}^{-1}\mathbb{V}\,]^{-1},
\label{eq:t-matrix}
\end{equation}
which, together with Eq.~(\ref{eq:cont}), determines the amplitudes $t_{j'j}$ and $r_{j'j}$ for the delta-potential in space.

\twocolumngrid

\end{document}